\documentclass[prl,floatfix,twocolumn,superscriptaddress,showpacs,preprintnumbers,longbibliography,nofootinbib]{revtex4-2}
\usepackage{color}
\usepackage[utf8]{inputenc}
\usepackage[usenames,dvipsnames]{xcolor}
\usepackage{amsmath}
\usepackage{amssymb}
\usepackage{graphicx}
\graphicspath{{graphics/}}
\usepackage{epsfig}
\usepackage{dcolumn}% Align table columns on decimal point
\usepackage{bm}% bold math
\usepackage{mathrsfs}
\usepackage{multirow}
\usepackage[all]{xy}
\usepackage{pbox}
\usepackage{lipsum}
\usepackage{verbatim}
\usepackage{braket}
\usepackage{dsfont}
\usepackage{array}
\usepackage{makecell}
\usepackage{tabularx}
\usepackage{isotope}
\usepackage{xr}
\usepackage{amsthm} 
\usepackage[normalem]{ulem} 
\usepackage{physics}
\usepackage{float}
\usepackage{placeins}

\usepackage[colorlinks=true,linkcolor=blue,urlcolor=blue,citecolor=blue,pdfusetitle]{hyperref}

\usepackage{tikz}
\newcommand{\tikzcircle}[2][red,fill=red]{\tikz[baseline=-0.5ex]\draw[#1,radius=#2] (0,0) circle ;}%

\begin{document}
\title{Non-linear classification capability of quantum neural networks due to emergent quantum metastability}
\author{Mario Boneberg}
\affiliation{Institut f\"ur Theoretische Physik and Center for Integrated Quantum Science and Technology, Universit\"at Tübingen, Auf der Morgenstelle 14, 72076 T\"ubingen, Germany}
\author{Federico Carollo}
\affiliation{Institut f\"ur Theoretische Physik and Center for Integrated Quantum Science and Technology, Universit\"at Tübingen, Auf der Morgenstelle 14, 72076 T\"ubingen, Germany}
\author{Igor Lesanovsky}
\affiliation{Institut f\"ur Theoretische Physik, Universit\"at Tübingen and Center for Integrated Quantum Science and Technology, Auf der Morgenstelle 14, 72076 T\"ubingen, Germany}
\affiliation{School of Physics and Astronomy and Centre for the Mathematics and Theoretical Physics of Quantum Non-Equilibrium Systems, The University of Nottingham, Nottingham, NG7 2RD, United Kingdom}

\begin{abstract}
The power and expressivity of deep classical neural networks can be attributed to non-linear
input-output relations. Such non-linearities are at the heart of many computational tasks, such as data classification and pattern recognition. Quantum neural networks, on the other hand, are necessarily linear as they process information via unitary operations. Here we show that effective non-linearities can be implemented in these platforms by exploiting the relationship between information processing and many-body quantum dynamics. The crucial point is that quantum many-body systems can show emergent collective behavior in the vicinity of phase transitions, which leads to an effectively non-linear dynamics in the thermodynamic limit. In the context of quantum neural networks, which are necessarily finite, this translates into metastability with transient non-ergodic behavior. By using a quantum neural network whose architecture is inspired by dissipative many-body quantum spin models, we show that this mechanism indeed allows to realize non-linear data classification, despite the underlying dynamics being local and linear. Our proof-of-principle study may pave the way for the systematic construction of quantum neural networks with emergent non-linear properties.
\end{abstract}

\maketitle

\noindent {\bf Introduction.---} Data processing and physical dynamics merely represent two different perspectives of the same phenomenon. An illustrative example is the \textit{Hopfield neural network} \cite{hopfield1982,amit1985a,amit1985b,amit1987}. It can be formulated in terms of spin degrees of freedom and allows to associate an input with previously stored patterns. This so-called memory retrieval is realized by implementing a discrete-time spin dynamics, that --- by minimizing a suitably chosen energy function --- takes an initial spin configuration to the nearest fixed point. In a machine learning scenario the aim is usually to infer this dynamical rule by supplying a data set and minimizing a suitable loss function under a given learning protocol \cite{bishop2006,nielsen2015, goodfellow2016}. Also the Hopfield model has been recently reinterpreted in such a context for supervised and unsupervised learning tasks \cite{alemanno2023, agliari2022}. 

Worldwide efforts are currently targeting the implementation of quantum versions of artificial neural networks \cite{schuld2014,dallaire2018,killoran2019, mangini2021,mujal2021, bravo2022, schuld2022}. Exploiting superposition, coherence and non-classical correlations such as entanglement, they promise to recognize 'atypical patterns' \cite{biamonte2017,cerezo2022} and seem to be promising candidates for achieving a quantum advantage on current so-called NISQ (noisy intermediate-scale quantum) hardware \cite{preskill2018,cerezo2021}. Quantum neural networks are necessarily linear. This may be a concern \cite{hornik1989,schuld2014,lecun2015}, because the expressive power of classical machine learning architectures --- with important exceptions, such as support vector machines \cite{cortes1995} --- draws substantially from non-linearities. However, as the Hopfield model shows, computational power and specifically the capability to retrieve patterns can be an emergent effect. This means it manifests as a limiting case of linear and ergodic dynamics.

Classical Hopfield neural networks and also  its dissipative quantum generalizations \cite{rotondo2018, fiorelli2022} retrieve patterns by dynamically approaching distinct stationary states (stable fixed points), depending on the basin of attraction within which the initial state is located \cite{lewenstein2021,bodeker2022}. This breaking of ergodicity, i.e. the formation of multiple distinct fixed points, is (in the absence of symmetries) only possible in the thermodynamic limit and is further amplified in the Hopfield model via all-to-all spin interactions. Here the dynamics becomes effectively non-linear, which can also be shown formally \cite{rotondo2018, fiorelli2022}.

\begin{figure}[t]
\centering
    \includegraphics[width=0.9\linewidth]{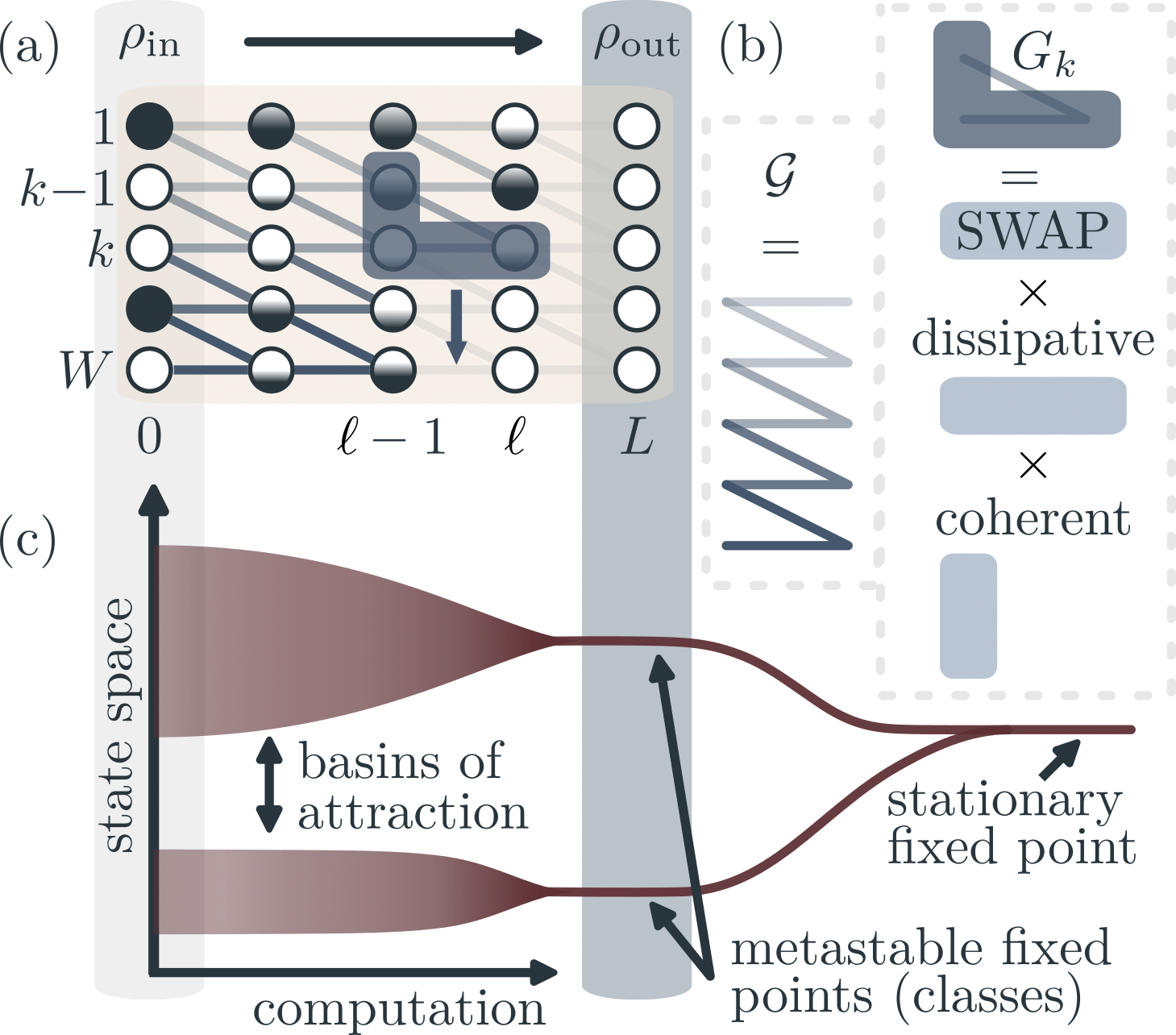}
    \caption{\textbf{Quantum neural network, perceptrons and metastability.} (a) The network consists of qubits forming a rectangular lattice of $L+1$ layers and width $W$. The qubits can be in state $\ket{0}$ (graphically represented by an empty circle $\tikzcircle[black, fill=white]{2pt}$), $\ket{1}$ (represented by $\tikzcircle[black, fill=black]{2pt}$). Partially filled circles indicate quantum properties such as e.g. superpositions.
    (b) Perceptrons are represented by local unitary gates $G_k$ (blue shapes/lines), which only act on two adjacent layers and are applied sequentially along a layer. They form the global gate $\mathcal{G}$ which is applied successively from left to right in order to compute an output state $\rho_{\mathrm{out}}$ on layer $L$ from an input state $\rho_{\mathrm{in}}$ on layer $0$. 
    (c) The repeated application of the global gate implements a dissipative spin dynamics which can show metastability: for intermediate times --- translating into a certain network depth $L$ --- ergodicity is approximately broken and clusters of initial states (basins of attraction) evolve towards different metastable fixed points. A deeper network would feature a unique output state (stationary fixed point). Choosing the depth $L$ of the quantum neural network such that it coincides with the metastable window (shaded region) allows to classify input states as they evolve to distinct output states.}
    \label{Fig1}
\end{figure}

In finite systems, such as a realistic quantum neural network [see sketch in Fig. \ref{Fig1}(a,b)], it is neither possible to approach the thermodynamic limit nor can one realize all-to-all connectivity amongst neurons. Nevertheless, effectively non-linear collective behavior may be exploited for data processing and classification: for example, remnants of ergodicity breaking manifest in finite systems in distinct metastable states, which are approximately stationary over long but transient timescales \cite{macieszczak2016,rose2016,macieszczak2021}.  During this transient, ergodicity is effectively broken and each metastable state possesses its own basin of attraction as sketched in Fig. \ref{Fig1}(c). This is a mechanism that enables the classification of data. 

Building on this idea, we put forward an approach to data classification in large-scale, locally-connected quantum feed-forward neural networks  \cite{beer2020, gillman2022b,beer2022, boneberg2023}. These networks are defined on a  qubit lattice which is organized into layers [cf. Fig. \ref{Fig1}(a,b)]. Sequences of local, translation-invariant gates are applied successively from one end to the other, operating as quantum perceptrons. As a consequence, the reduced state on the input layer is propagated through hidden layers, towards the output layer, via a quantum channel \cite{nielsen2010}, 
thereby implementing a discrete-time dissipative quantum many-body spin dynamics.
As a proof-of-principle demonstration we employ a network whose perceptrons are inspired by the dissipative quantum Ising model \cite{ates2012,weimer2015,rose2016,letscher2017,jiasen2018}. This model hosts emergent metastable fixed points, associated with phase coexistence near a first-order phase transition or crossover. We simulate the ensuing quantum dynamics on classical hardware using tensor network methods \cite{orus2014,paeckel2019,gillman2020,gillman2021a,gillman2021b,gillman2022a,gillman2022a, gillman2022b} and show that emergent collective phenomena indeed enable our neural network to separate inputs into two classes [see Fig. \ref{Fig1}(c)]. 
\\

\noindent {\bf Quantum neural network architecture and link to dissipative spin dynamics.---} The quantum neural network is defined on a two-dimensional lattice with $L + 1$ layers and width $W$ [see Fig. \ref{Fig1}(a)] \cite{lesanovsky2019,gillman2020,beer2020,gillman2021a,gillman2021b,gillman2022a,gillman2022b,boneberg2023}. Each lattice site contains a qubit with basis states $\ket{0}$ (vacuum) and $\ket{1}$, representing a node of the network. Initially, all layers apart from the first one are in the vacuum state. The first layer contains the input state. The perceptrons of the quantum neural network are given by quantum gates $G_{k}$ which act between two adjacent layers and are localized in a neighborhood of a site $k$ as illustrated in Fig. \ref{Fig1}(a,b). They are applied sequentially along a layer in order to form the global gate $\mathcal{G}= \prod_{k=1}^W G_{k}$ (up to boundary terms, see supplemental material \cite{SM}), as shown in Fig. \ref{Fig1}(b). The successive application of the global gate from left to right propagates the reduced state on the first layer, $\rho_{\mathrm{in}}$ through the hidden layers of the network to the final layer, yielding $\rho_{\mathrm{out}}$ \cite{beer2020,beer2022}. The horizontal axis can thus be interpreted as a time axis with each layer of the network representing an instance of a many-body system consisting of $W$ spins.

In the following we elaborate on this idea by considering the evolution among two adjacent layers. The reduced state on layer $\ell$, $\rho ( \ell)$, follows the recurrence relation \cite{gillman2022b,boneberg2023}
\begin{equation} \label{recurrence}
    \rho (\ell) = \Tr_{\ell-1}\left(\mathcal{G} \rho(\ell-1) \otimes \outerproduct{\mathbf{0}}{\mathbf{0}}   \mathcal{G}^\dagger \right).
\end{equation}
Here, the trace is over all sites in layer $\ell-1$ and $\outerproduct{\mathbf{0}}{\mathbf{0}}$ is the vacuum state on layer $\ell$. Note, that in what follows operators on the left and right of a Kronecker product act on layers $\ell-1$ and $\ell$, respectively.
The local gate, which links sites of adjacent layers is parametrized as \cite{boneberg2023}
\begin{equation} \label{localgate}
    G_{k} = \mathrm{SWAP}_{k} \times e^{-i \sqrt{\delta t} \big( J_{k} \otimes \sigma_{k}^+ + \text{h.c.}\big)} \times e^{-i \delta t H_{k} \otimes \mathds{1}},
\end{equation}
with a graphical representation shown in Fig.~\ref{Fig1}(b). It consists of three parts, which have a transparent interpretation: the rightmost term acts non-trivially on layer $\ell-1$ where it generates a coherent evolution under the Hamiltonian $H_k$ over the time step $\delta t$. The middle term connects adjacent layers and generates, as we will see shortly, dissipative dynamics governed by the so-called jump operator $J_k$. Note, that the operator $\sigma^+ = \sigma^x + i \sigma^y$, which acts on layer $\ell$ is defined via the Pauli matrices $\sigma^\alpha, \alpha =x,y,z$. Finally, the $\mathrm{SWAP}_k$ operator on the left interchanges the state of the qubits located on sites $(k,\ell-1)$ and $(k,\ell)$ and ultimately leads to the propagation through the network. This interpretation becomes immediately apparent when expanding Eq. (\ref{recurrence}) to first order in $\delta t \ll 1$, which yields a Markovian open system quantum dynamics: $\rho(\ell) \approx \rho (\ell-1) + \delta t \mathcal{L}[\rho(\ell-1)] \approx e^{\delta t  \mathcal{L}}[\rho(\ell-1)]$ \cite{boneberg2023,lorenzo2017,ciccarello2017,ciccarello2022,cattaneo2021,cattaneo2022} with the \textit{Lindblad generator} \cite{lindblad1976,gorini1976,breuer2002}
\begin{align*}
    \mathcal{L}[\cdot] =& -i\left[\sum_{k=1}^W H_{k}, \cdot \right] + \sum_{k=1}^W \left( J_{k} \left( \cdot \right) J_{k}^\dagger - \frac{1}{2} \left\{ J_{k}^{\dagger} J_{k}, \cdot \right\} \right) .
\end{align*}
\\
This equation gives a transparent interpretation for the emergence of effective non-linearities in quantum neural networks. In the thermodynamic limit, i.e. when the network width $W \to \infty$, its stationary state can feature phase transitions which are accompanied by emergent collective behavior, such as non-analytic points and ergodicity breaking. The latter may manifest at finite network width $W$ through metastability.
To illustrate this we choose a form of the local gate (\ref{localgate}), which is inspired by the \textit{dissipative quantum Ising model} \cite{ates2012,weimer2015,rose2016,jiasen2018}: $H_k = \frac{\Omega}{2} \sigma_k^x + \frac{V}{4} \sigma_{k-1}^z \sigma_k^z$ and $J_k = \sqrt{\kappa} \sigma_k^-$, with $\Omega, V$ and $\kappa$ being tunable parameters, which can be interpreted as transverse field, spin-spin interaction strength and dissipation rate, respectively. Due to translation invariance of $H_k$ and $J_k$ also the local gate \eqref{localgate} is translation invariant. One reason for considering this model is that it features a crossover or phase transition, depending on the dimension, associated with the breaking of a $\mathbb{Z}_2$-symmetry which is not manifest at the microscopic level \cite{marcuzzi2014,overbeck2017}. This means that all collective effects, such as ergodicity breaking are emergent and do not require the presence of an explicit symmetry.

\begin{figure}[t]
\centering
    \includegraphics[width=\linewidth]{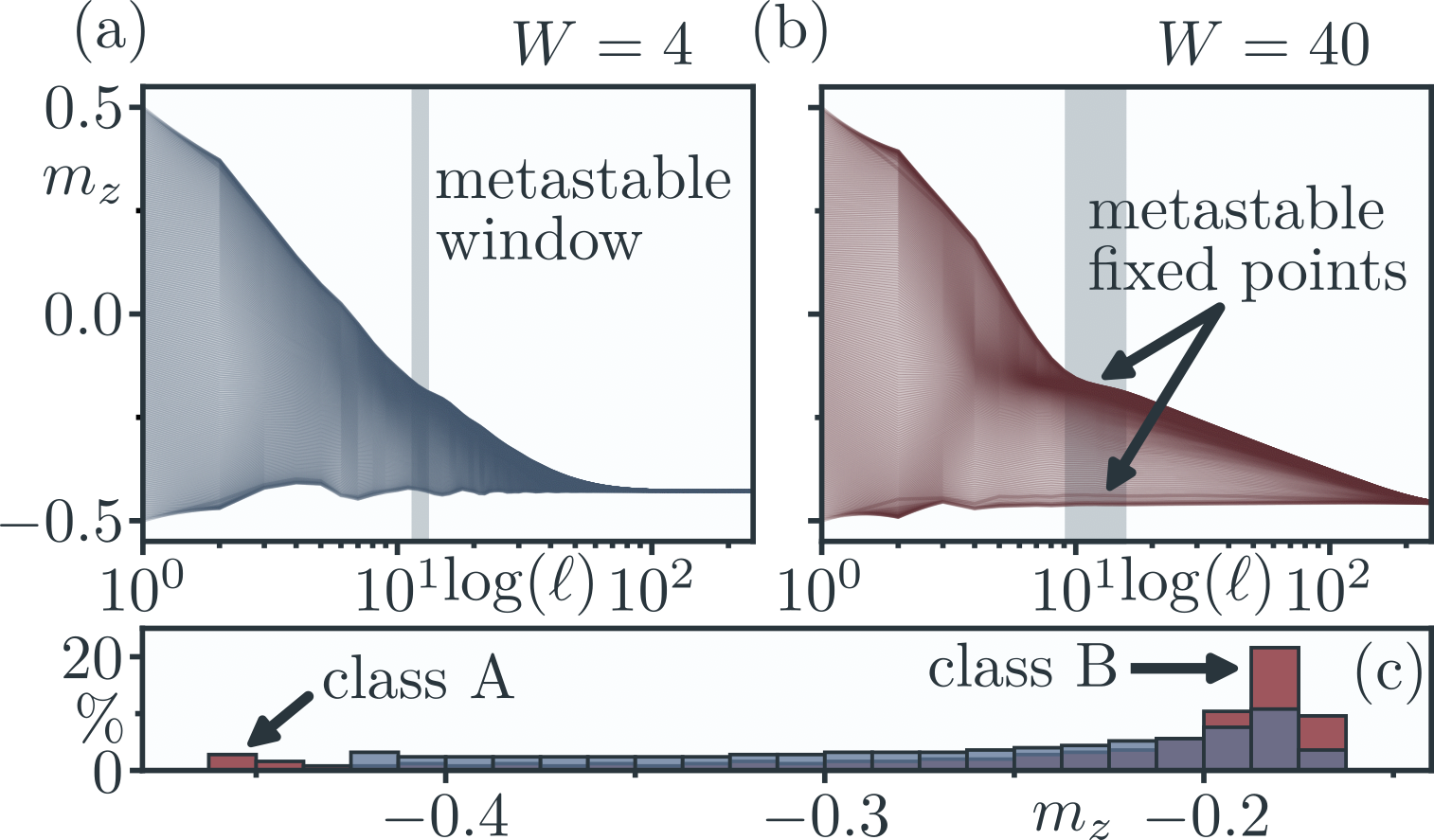}
    \caption{\textbf{Emergent metastability and data classification.} (a) Evolution of the magnetization order parameter $m_z$ of the reduced state $\rho(\ell)$ [Eq. (\ref{recurrence})]. To generate the plot, we chose $200$ initial product states whose magnetization $m_z$ covers the interval $[-0.5, 0.5]$ uniformly (see main text for details). The local gates \eqref{localgate} are chosen such that they correspond to the dissipative quantum Ising model (open boundary conditions), with parameters $\Omega = 59 \kappa , V = 250 \kappa, \kappa \delta t=0.1$. To generate the data we employed matrix product states with bond dimension $\chi =120$ (see supplemental material \cite{SM} for details). For the chosen network width ($W=4$) only a small metastability window (shaded region) occurs. (b) As the width is increased to $W=40$, the size of the metastability window, which features two metastable fixed points, increases significantly. (c) Histogram of the magnetization values on layer $\ell=11$, which is located in the metastability window. The data corresponding to panel (a) and (b) is shown in blue and red, respectively. The distribution is bimodal with peaks located at magnetization values that correspond to the two metastable fixed points. The peaks can interpreted as different classes into which the initial states fall, here denoted as class $A$ and $B$. Note, that as the network width $W$ increases, the peaks are expected to become more pronounced. This means that the background of initial states, which are not classified, i.e. whose final magnetization values are located in between the peaks, is expected to shrink. Also note, that the peaks are not expected to be of equal height, since the basins of attraction of the dissipative Ising model are asymmetric, see Ref. \cite{rose2016}.}
    \label{Fig2}
\end{figure}

In Fig.~\ref{Fig2} we show that with this choice of the gate (\ref{localgate}), the network indeed features metastable fixed points as illustrated in Fig. \ref{Fig1}(c). To do so, we analyze the quantity
\begin{equation*}
    m_z  = \Tr(\frac{1}{2W}\sum_{k=1}^W \sigma_k^z\,  \rho),
\end{equation*}
which in the context of the Ising model can be interpreted as the magnetization order parameter. To generate the figure we chose initial pure states $\bigotimes_{k=1}^W \outerproduct{\psi}{\psi}$ with $\ket{\psi}=\cos(\theta/2) \ket{0} + \sin(\theta/2) \ket{1}$, $\theta = \arccos(2m_z)$, where the magnetizations $m_z$ cover the interval $[-0.5, 0.5]$. We follow the evolution of $m_z$ through the hidden layers. For intermediate depth we see that the magnetization bunches around two metastable states, reflected in a bimodal distribution. Physically this originates from the coexistence of a paramagnetic and a ferromagnetic state near a first-order phase transition \cite{ates2012,marcuzzi2014,rose2016}. For larger depth the dynamics approaches a unique stationary fixed point. Comparing panels (a) and (b) one observes that the metastable window increases with network depth $W$ and the bimodal character of the distribution is enhanced [panel (c)]. These two modes can then be utilized for separating initial states into two classes, here referred to as class A and B.

\begin{figure*}[t]
\centering
    \includegraphics[width=\textwidth]{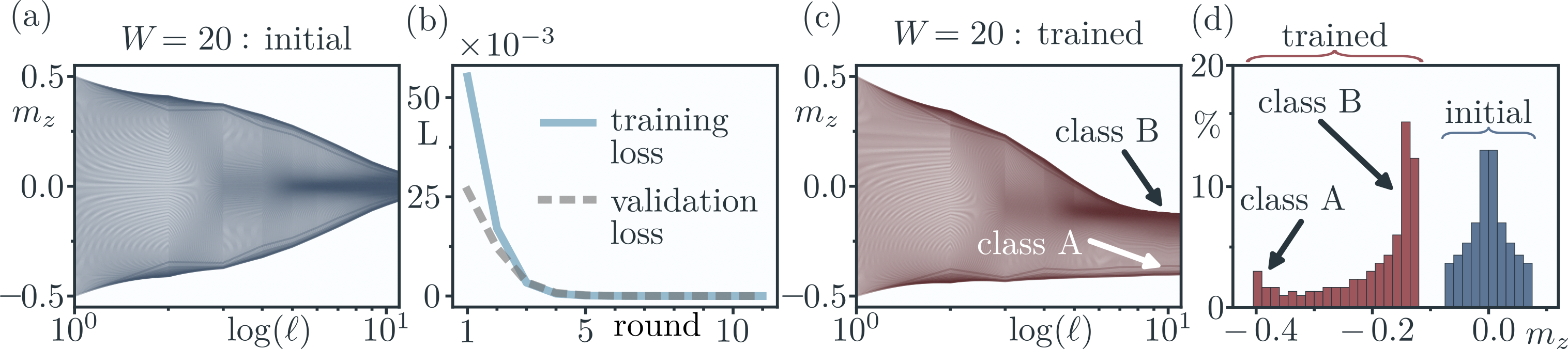}
    \caption{\textbf{Quantum neural network training.} (a) Evolution of the magnetization order parameter $m_z$ (c.f. Fig. \ref{Fig2}) for the untrained network. There is no metastable behavior and the distribution of output magnetizations is unimodal as shown in panel (d). (b) Training loss (blue) and validation loss (grey, dashed) for $11$ training rounds; both decrease to zero. 
    In our protocol we chose the learning rate $\epsilon=10$ and a matrix product state bond dimension $\chi=96$. (c) Evolution of the magnetization $m_z$ in the trained network showing metastability which allows to identify two classes ($A$, $B$) of initial states. (d) Histogram for trained network (red) and initial network (blue).}
    \label{Fig3}
\end{figure*}

\noindent {\bf Training protocol.---} To accomplish a specific task such as classification, the network needs to be trained. The expectation is that this training will set the parameters of the local gate (\ref{localgate}) such that metastable fixed points emerge. This is indeed the case, as we will show later, but first we briefly outline the training procedure, which follows closely the idea of Ref. \cite{beer2020}. Full details are found in the supplemental material \cite{SM}.

The training data set consists of $P$ pairs of the form $\Big(\rho_{\mathrm{in}}^{(x)}, m_{z,\mathrm{out}}^{(x)}\Big)$, where $x=1,\ldots,P$. Here $\rho_{\mathrm{in}}^{(x)}$ is an input state and $m_{z,\mathrm{out}}^{(x)}$ the associated output magnetization. We use the loss function  
\begin{equation} \label{loss}
    L = \frac{1}{P} \sum_{x=1}^P \Tr(\bigg( \frac{1}{2W} \sum_{k=1}^W \sigma_k^z  - m_{z,\mathrm{out}}^{(x)} \bigg) \rho_{\mathrm{out}}^{(x)})^2 ,
\end{equation}
which quantifies the average, over all training pairs, of the squared differences between desired and actual output magnetization. 
An advantage of this loss function is that it does not rely on quantum state overlaps, which become exponentially small with growing network width $W$. Furthermore, fluctuations of $L$ decrease with increasing $W$ due to the law of large numbers. This may help to avoid so-called barren plateaus \cite{mcclean2018,sharma2022,cerezo2023}. To minimize $L$ during the training, we update the local gates $G_k$ [Eq. \eqref{localgate}] repeatedly via multiplication by unitary update gates of the form $e^{-i \epsilon K}$. Here $\epsilon$ is a scalar, called the learning rate, and $K$ is the update-parameter matrix (see \cite{beer2020,SM} for details) which needs to be chosen such that one moves into the direction of the fastest decrease of the loss function \cite{beer2020}. This is accomplished by computing the change of the loss function under the update $G_k \rightarrow G_k'$ to first, i.e. linear, order in $\epsilon$. The direction of the negative gradient then yields analytical expressions for the matrix $K$. These updates depend on the one hand on the reduced states obtained by propagating the training states via the quantum channel \eqref{recurrence}. On the other hand they depend on the reduced states of the layers obtained by backpropagating 'error operators' for the training magnetizations via the adjoint of the channel of the map \eqref{recurrence}. To implement this algorithm on a classical computer for large widths $W$, we necessarily have to implement tensor network methods \cite{SM,gillman2021b,paeckel2019}. 

\noindent {\bf Training results.---} 
In the following we present a proof-of-principle demonstration of the training. Our aim here is not to actually solve a practically relevant classification task. Instead we will show that an initially trivial network acquires emergent metastable behavior and data classification capability through training. One reason for resorting to such simplified scenario is that the implementation of the above-mentioned training algorithm via tensor networks on classical computers is highly computationally demanding for the rather large network sizes (hundreds of qubit) that we consider.  
The training data is generated using the dissipative Ising model parametrization of the gate (\ref{localgate}) with parameters $\Omega=70  \kappa, V=250  \kappa, \kappa \delta t =0.1$ and on a network containing $200$ qubits: width $W=20$ and $L=10$. We generate the data for input states, $\rho_{\mathrm{in}}^{(x)}$, with magnetization in the interval $[-0.5,0.5]$ and calculate the magnetization $m_{z,\mathrm{out}}^{(x)}$ at the output. For these network parameters, metastable behavior results in a bimodal distribution of the output magnetizations, c.f. Fig. \ref{Fig2}(b), and we may identify two classes $A$ and $B$.

With this data we train a  network whose local gates are initially expressed in terms of local Ising Hamiltonians $H_k$, but with jump operators $J_k= -i \sqrt{\kappa} \sigma_k^y$. As illustrated in Fig~\ref{Fig3}(a), this network features no metastability and the distribution of the output magnetization is unimodal. During training the local gate \eqref{localgate} is updated via
\begin{align*}
    e^{-i \sqrt{\delta t} V_{k}} &\rightarrow e^{-i\epsilon\frac{  \sqrt{\delta t}}{2} \tilde{V}_{k}} e^{- i \sqrt{\delta t}  V_{k}} e^{-i \epsilon\frac{ \sqrt{\delta t}}{2} \tilde{V}_{k}}  \\
    e^{-i \delta t H_{k} \otimes \mathds{1}} &\rightarrow e^{-i \delta t H_{k} \otimes \mathds{1}}  e^{-i \epsilon \delta t \tilde{H}_{k} \otimes \mathds{1}}, 
\end{align*}
where $V_{k} = J_{k} \otimes \sigma_{k}^+ + \mathrm{h.c.}$ The operators $\frac{\sqrt{\delta t}}{2}\tilde{V}_{k} = \tilde{J}_{k} \otimes \sigma_{k}^+ + \mathrm{ h.c.}$ and $\delta t \tilde{H}_k$ represent here the update-parameter matrix $K$, which was previously mentioned. This scheme ensures that to first order in $\delta t$, the network dynamics remains governed by a Lindblad master equation, where the Hamiltonian and jump operators are updated as $H_{k}\rightarrow H_{k} + \epsilon \tilde{H}_{k}$ and $J_{k}\rightarrow J_{k} + \epsilon \tilde{J}_{k}$. The updates are chosen such that translation invariance is preserved. In the current example, we make a further simplification: we do not update the Hamiltonian, i.e. $\tilde{H}_{k}=0$, but solely the jump operator via a shift of the form $\tilde{J}_{k} \propto \sigma_{k}^x$ (for details, see \cite{SM}). 

In Fig.~\ref{Fig3}(b) we display the evolution of the loss \eqref{loss} during training as solid blue line. It approaches zero, indicating that the network learns, on average, to correctly assign the output to the input of each training pair. The trained quantum neural network should, furthermore, be able to predict the output magnetizations for input states which have not been used for training. This capability is quantified by the validation loss, which is computed via Eq.~\eqref{loss} using $10$ validation data pairs. In Fig.~\ref{Fig3}(b) we depict this quantity as grey, dashed line, showing that also the validation loss approaches zero. The dynamics of the magnetization in the trained network is shown in Fig.~\ref{Fig3}(c). A comparison with Fig.~\ref{Fig3}(a) clearly evidences metastable behavior in the vicinity of the output layer. As can be seen in panel (d), this is reflected in a bimodal distribution of the magnetization at the output, which allows to assign input states to the two classes $A$ and $B$. 

In this example, effectively only one parameter of the gate \eqref{localgate} is varied. This restriction is due to the immense computational cost that is connected to the training of networks, with the sizes considered here, on classical computers. In this simple setting, the training loss is well-behaved and quickly approaches zero. However, in more general scenarios the loss function landscape in parameter space may be governed by local minima in which the algorithm might get stuck. This may lead to a stagnation of the loss at finite values and/or a bad generalization beyond the training data. Local minima which cause stagnation of the loss may be escaped by incorporating randomness and information from previous steps (momentum) into the update \cite{nielsen2015,goodfellow2016,kingma2014,ruder2016,bondarenko2020}. 

\noindent {\bf Summary and outlook.---} We have shown that information propagation in quantum neural networks is intimately connected with the dynamics of dissipative quantum spin systems. This opens opportunities for exploiting emergent many-body phenomena for enhancing the computational capabilities of these platforms. We have illustrated this in a proof-of-principle demonstration of a network that has the capability to classify input data by exploiting quantum metastability. A crucial point is that these emergent collective effects are only present in large quantum neural networks. Their simulation on classical computers is inefficient, which also posed severe limitations for our theoretical study. However, the ultimate goal is to realize these networks on scalable quantum hardware. Given the effectively dissipative nature of the quantum neural network dynamics a certain amount of dissipation on these devices might even be tolerable, assuming that emergent collective features are not qualitatively altered. This is an intriguing prospect, which suggests that certain quantum machine learning tasks may already be implementable on NISQ devices, which offer large qubit numbers.

\begin{acknowledgements}
\noindent \textbf{Acknowledgments.} 
The research leading to these results has received funding from the Deutsche Forschungsgemeinschaft (DFG, German Research Foundation) under Project No. 449905436, as well as through the Research Unit FOR 5413/1, Grant No. 465199066. This Project has also received funding from the European Union’s Horizon Europe research and innovation program under Grant Agreement No. 101046968 (BRISQ), and from EPSRC under Grant No. EP/V031201/1. FC is indebted to the Baden-Württemberg Stiftung for the financial support of this research Project by the Eliteprogramme for Postdocs. IL is a member of the Machine Learning Cluster of Excellence, funded by the Deutsche Forschungsgemeinschaft (DFG, German Research Foundation) under Germany’s Excellence Strategy—EXC Number 2064/1 - Project Number 390727645. The authors acknowledge support by the state of Baden-Württemberg through bwHPC and the German Research Foundation (DFG) through grant no INST 40/575-1 FUGG (JUSTUS 2 cluster).
\end{acknowledgements}

\bibliography{Reference}

\setcounter{equation}{0}
\setcounter{figure}{0}
\setcounter{table}{0}
\makeatletter
\renewcommand{\theequation}{S\arabic{equation}}
\renewcommand{\thefigure}{S\arabic{figure}}

\makeatletter
\renewcommand{\theequation}{S\arabic{equation}}
\renewcommand{\thefigure}{S\arabic{figure}}

\onecolumngrid
\newpage

\begin{center}
{\Large SUPPLEMENTAL MATERIAL}
\setcounter{page}{1}
\end{center}
\begin{center}
\vspace{0.8cm}
{\Large Non-linear classification capability of quantum neural networks due to emergent quantum metastability}
\end{center}
\begin{center}
Mario Boneberg,$^{1}$ Federico Carollo$^1$ and Igor Lesanovsky$^{2,3}$
\end{center}
\begin{center}
$^1${\em Institut f\"ur Theoretische Physik and Center for Integrated Quantum Science and Technology,}\\
{\em Universit\"at Tübingen, Auf der Morgenstelle 14, 72076 T\"ubingen, Germany}\\
$^2$ {\em Institut f\"ur Theoretische Physik, Universit\"at Tübingen and Center for Integrated Quantum Science and Technology,}\\
{\em  Auf der Morgenstelle 14, 72076 T\"ubingen, Germany}\\
$^3$ {\em School of Physics and Astronomy and Centre for the Mathematics}\\
{\em  and Theoretical Physics of Quantum Non-Equilibrium Systems,}\\
{\em  The University of Nottingham, Nottingham, NG7 2RD, United Kingdom}
\end{center}

\allowdisplaybreaks

Here we present in more detail the classical learning algorithm for our quantum neural networks (QNNs) and explain how tensor network techniques can be utilized to simulate it for large width $W$. We focus on the special case used in training based on data from the open Ising model. We structure this as follows. First we specify the form of our local unitary gates (perceptrons) and the channel our network computes \cite{boneberg2023}. Then we define the update protocol decreasing the loss function; this follows closely the strategy of Ref.~\cite{beer2020}. For large $W$, classically computing the updates is costly. To deal with such cases we introduce techniques based on matrix-product states (MPSs) and matrix-product operators (MPOs). Finally, we discuss how metastability can occur in Markovian open quantum dynamics.

\section{Perceptrons and output channel}
For a certain input state $\rho_0^{(x)}$ on layer $0$, labeled by $x$, the corresponding output state of our QNN on layer $L$ is computed as
\begin{align} \label{input-output}
    \rho_{L}^{(x)}(s) =& \Tr_{\neq L}\Bigg( \prod_{\ell = L}^{1} \mathcal{G}_{\ell}(s)   \Bigg[ \rho_0^{(x)} \otimes \bigotimes_{i = 1}^{L} \ket{\mathbf{0}}\bra{\mathbf{0}} \Bigg] \prod_{\ell = 1}^{L}  \mathcal{G}_{\ell}^{ \dagger}(s)  \Bigg) ,
\end{align}
where $\ket{\mathbf{0}}\bra{\mathbf{0}}$ is the vacuum state on one layer and $s$ indicates the current parameterization of the QNN. The $\Tr_{\neq L}$ refers to the trace over all layers but the $L$th. Note that here for a state $\rho$ we write the layer ${\ell}$ it is defined on as an index. In addition, we label the global gate $\mathcal{G}$ by the layer $\ell$, to indicate that it represents an embedded operator acting nontrivially on layers $\ell -1, \ell$. It is defined as \cite{boneberg2023}
\begin{align*}
    \mathcal{G}_{\ell} &=  \mathrm{SWAP}_{1,\ell}  \left( \prod_{k=2}^{W} \mathrm{SWAP}_{k,\ell}  e^{-i \sqrt{\delta t} V_{k,\ell}}  e^{-i \delta t H_{k,\ell}}   \right)     e^{-i \sqrt{\delta t} V_{1,\ell}}  e^{-i \delta t H_{1,\ell}}   =  L_{1,\ell}  \left( \prod_{k=2}^{W}   G_{k,\ell}    \right)    R_{1,\ell}   ,
\end{align*}
with the local gates $L_{1,\ell},G_{k,\ell},R_{1,\ell}$. For $\vec{\alpha}=(\alpha_1,\alpha_2)^T$, $\alpha_1, \alpha_2 = 0,1,2,3$, $c^{\vec{\alpha} }\in \mathbb{C},d^{\vec{\alpha} }\in \mathbb{R}$ and $\sigma^-= \left( \sigma^{+} \right)^\dagger$ we have
\begin{align*}
    H_{k, \ell} &= \sum_{\vec{\alpha}} d_k^{\vec{\alpha}} \sigma_{k-1, \ell-1}^{\alpha_1}  \sigma^{\alpha_2}_{k, \ell-1} = \sum_{\vec{\alpha}} d_k^{\vec{\alpha}} \hat{S}^{\vec{\alpha}}_{k,\ell}  \\
    V_{k, \ell} &=   J_{k,\ell}  \sigma^+_{k,\ell} + \Big( J_{k,\ell} \Big)^\dagger  \sigma^-_{k, \ell}  = \sum_{\vec{\alpha}} \Big(  c_k^{\vec{\alpha}} \sigma_{k-1, \ell-1}^{\alpha_1}  \sigma^{\alpha_2}_{k, \ell-1}  \sigma^+_{k,\ell} + \left(c_k^{\vec{\alpha}} \right)^* \sigma_{k-1, \ell-1}^{\alpha_1}  \sigma^{\alpha_2}_{k, \ell-1}  \sigma^-_{k, \ell} \Big) \\
    &= \sum_{\vec{\alpha}} \Big(  c_k^{\vec{\alpha}} \hat{S}^{+,\vec{\alpha}}_{k,\ell} + \left(c_k^{\vec{\alpha}} \right)^* \hat{S}^{-,\vec{\alpha}}_{k,\ell} \Big) \\
    \mathrm{SWAP}_{k, \ell} &= \sum_{\alpha =0,1,2,3} \frac{1}{2} \Big( \sigma^{\alpha}_{k, \ell-1} \otimes  \sigma^\alpha_{k,\ell}  \Big) \\
        c_k^{\vec{\alpha}} &= \chi_{k \geq 2} c^{\vec{\alpha}} + \chi_{k = 1} \delta^{\alpha_10} c^{\vec{\alpha}} \\
     d_k^{\vec{\alpha}} &= \chi_{k \geq 2} d^{\vec{\alpha}} + \chi_{k = 1} \delta^{\alpha_10} d^{\vec{\alpha}} ,
\end{align*}
where $\chi_{\tilde{k} \geq 2}$ is one if $\tilde{k} \geq 2$ and zero otherwise. On the other hand, $\chi_{\tilde{k} = 1}$ is one if $\tilde{k} = 1$ and zero otherwise. Here we label also the Hamiltonian $H_k$, the jump operator $J_k$ (and $V_k$) (see main text) by $\ell$ to indicate that they represent embedded operators; $H_{k,\ell}$ and $J_{k,\ell}$ both act nontrivially only on sites $k-1,k$ of layer $\ell$. Analogously, the operator $\mathrm{SWAP}_{k, \ell}$ interchanges the roles of sites $k$ on layers $\ell-1,\ell$. For a Pauli matrix a subscript $k,\ell$ indicates that it acts nontrivially only on site $k$ of layer $\ell$. With the last two equations we choose the gate $G_{k,\ell}$ translation invariant and implement open boundary conditions.\\
As explained in the main text and shown in Ref.~\cite{boneberg2023}, the output state can equivalently to Eq.~\eqref{input-output} be computed through the recurrence relation 
\begin{equation} \label{recurrence_sup}
    \rho_\ell^{(x)}(s) = \Tr_{\ell-1}\left(\mathcal{G}_{\ell}(s) \rho_{\ell-1}^{(x)}(s) \otimes \ket{\mathbf{0}}\bra{\mathbf{0}}   \mathcal{G}_{\ell}^\dagger(s) \right) 
\end{equation}
for the reduced states $\rho_\ell^{(x)}(s)$ of layer $\ell$. Importantly, for small $\delta t$ we may approximate
\begin{equation*}
\rho_\ell^{(x)}(s) \approx \rho_{\ell-1}^{(x)}(s) + \delta t \mathcal{L}(s)\left[\rho_{\ell-1}^{(x)}(s)\right] \approx e^{\delta t \mathcal{L}(s)}\left[\rho_{\ell-1}^{(x)}(s)\right] ,
\end{equation*}
and in the Lindblad generator $\mathcal{L}$ (see main text) the local and translation invariant Hamiltonians and jump operators have support on two sites.
\section{Training algorithm}
We train the QNN by repeatedly applying parametrized update unitaries to the local gates in order to decrease the loss function. In this section we explain the concrete update scheme and show how to choose the parameters in the update gates for fastest decrease of the loss.
\subsection{Update scheme}
As motivated in the main text, the QNN should still compute a Lindblad dynamics to first order in $\delta t$ after the update. This is achieved by 
\begin{align}
        e^{-i \sqrt{\delta t} V_{k, \ell}(s)} &\longrightarrow e^{-i \frac{\sqrt{\delta t}}{2} \epsilon \tilde{V}_{k, \ell}(s)} e^{-i \sqrt{\delta t} V_{k, \ell}(s)} e^{-i \frac{\sqrt{\delta t}}{2} \epsilon \tilde{V}_{k, \ell}(s)}   \label{update_1} \\  
        e^{-i \delta t H_{k, \ell}(s)} &\longrightarrow e^{-i \delta t H_{k, \ell}(s)} e^{-i \delta t \epsilon \tilde{H}_{k, \ell}(s)}  \label{update_2},
\end{align}
i.e. multiplication of update gates, with 
\begin{align*}
    \tilde{H}_{k, \ell} (s) &= \sum_{\vec{\alpha}} \tilde{d}_k^{\vec{\alpha}}(s) \hat{S}^{\vec{\alpha}}_{k,\ell}   \\
    \tilde{V}_{k, \ell}(s) &= \sum_{\vec{\alpha}} \Big(  \tilde{c}_k^{\vec{\alpha}}(s) \hat{S}^{+, \vec{\alpha}}_{k,\ell}  + \left(\tilde{c}_k^{\vec{\alpha}} \right)^* (s) \hat{S}^{-,\vec{\alpha}}_{k,\ell}  \Big) \\
    \tilde{c}_k^{\vec{\alpha}} &= \chi_{k \geq 2} \tilde{c}^{\vec{\alpha}} + \chi_{k = 1} \delta^{\alpha_10} \tilde{c}^{\vec{\alpha}} \\
     d_k^{\vec{\alpha}} &= \chi_{k \geq 2} \tilde{d}^{\vec{\alpha}} + \chi_{k = 1} \delta^{\alpha_10} \tilde{d}^{\vec{\alpha}},
\end{align*}
which preserves open boundary conditions and translation invariance of the gates. Indeed, to first order in $\delta t$ we have
\begin{align*}
    &e^{-i \frac{\sqrt{\delta t}}{2} \epsilon \tilde{V}_{k, \ell}(s)} e^{-i \sqrt{\delta t} V_{k, \ell}(s)} e^{-i \frac{\sqrt{\delta t}}{2} \epsilon \tilde{V}_{k,  \ell}(s)}  \\
    &= \mathds{1} -i \sqrt{\delta t} \left( V_{k, \ell}(s) + \epsilon \tilde{V}_{k, \ell}(s)  \right) - \frac{\delta t}{2} \left(  {V_{k, \ell}(s)} +  \epsilon {\tilde{V}_{k, \ell}(s)} \right)^2 \approx e^{-i \sqrt{\delta t} \left( V_{k, \ell}(s) + \epsilon \tilde{V}_{k, \ell}(s)  \right)}  = e^{- i \sqrt{\delta t} V_{k, \ell}(s+\epsilon)}
\end{align*}
and
\begin{align*}
    e^{-i \delta t H_{k, \ell}(s)} e^{-i \delta t \epsilon \tilde{H}_{k, \ell}(s)}  \approx \mathds{1} -i \delta t \Big( H_{k, \ell}(s) + \epsilon \tilde{H}_{k, \ell}(s) \Big) \approx e^{-i \delta t \big( H_{k, \ell}(s) + \epsilon \tilde{H}_{k, \ell}(s) \big)} = e^{-i \delta t H_{k, \ell}(s+\epsilon)} ,
\end{align*}
showing that also the form of the local gates is preserved within this approximation. As a consequence, the update effectively only shifts the parameters via
\begin{align*}
    V_{k, \ell}(s+\epsilon)= V_{k, \ell}(s) + \epsilon \tilde{V}_{k, \ell}(s) =&  \sum_{\Vec{\alpha}} \Big( c_k^{\Vec{\alpha}}(s) \hat{S}^{+,\vec{\alpha}}_{k,\ell}  +  \left(c_k^{\Vec{\alpha}} \right)^*(s) \hat{S}^{-,\vec{\alpha}}_{k,\ell}  \Big)  + \epsilon   \sum_{\Vec{\alpha}} \Big( \tilde{c}_k^{\Vec{\alpha}}(s) \hat{S}^{+,\vec{\alpha}}_{k,\ell}  +  \left(\tilde{c}_k^{\Vec{\alpha}} \right)^* (s) \hat{S}^{-,\vec{\alpha}}_{k,\ell}  \Big) \\
    =& \sum_{\Vec{\alpha}} \Big( \Big( c_k^{\Vec{\alpha}}(s) + \epsilon \tilde{c}_k^{\Vec{\alpha}}(s) \Big) \hat{S}^{+,\vec{\alpha}}_{k,\ell}   + \Big( c_k^{\Vec{\alpha}} (s) + \epsilon \tilde{c}_k^{\Vec{\alpha}} (s) \Big)^* \hat{S}^{-\vec{\alpha}}_{k,\ell}  \Big) \\
    =&  \left( J_{k,\ell}(s) + \epsilon \tilde{J}_{k,\ell}(s) \right)  \sigma^+_{k,\ell} + \left( J_{k,\ell}(s) + \epsilon \tilde{J}_{k,\ell}(s)  \right)^\dagger  \sigma^-_{k, \ell},
\end{align*}
and
\begin{align*}
    H_{k, \ell}(s+\epsilon) = H_{k, \ell}(s) + \epsilon \tilde{H}_{k, \ell}(s) 
    =& \sum_{\vec{\alpha}} d_k^{\vec{\alpha}}(s) \hat{S}^{\vec{\alpha}}_{k,\ell}  + \epsilon \sum_{\vec{\alpha}} \tilde{d}_k^{\vec{\alpha}}(s) \hat{S}^{\vec{\alpha}}_{k,\ell} 
    = \sum_{\vec{\alpha}} \Big( d_k^{\vec{\alpha}}(s) + \epsilon \tilde{d}_k^{\vec{\alpha}}(s) \Big) \hat{S}^{\vec{\alpha}}_{k,\ell} .
\end{align*}

\subsection{Update parameters}
So far, we only fixed the form of the update. In the following we explain how to choose the update parameters $\tilde{c}^{\vec{\alpha}},\tilde{d}^{\vec{\alpha}}$ such that the loss function is minimized the fastest. For this we need to compute the change of the loss 
\begin{equation*}
    \frac{dL}{ds} = \lim_{\epsilon \to 0} \frac{L(s+\epsilon) - L(s)}{\epsilon},
\end{equation*}
where $s+\epsilon$ indicates the parameterization of the QNN after the update. This is done by first approximating the updated output states $\rho_{L}^{(x)}(s + \epsilon )$ to first order. We define 
\begin{align*}
    A_{\tilde{k}, \tilde{\ell}} &=  \Bigg( \prod_{\ell=L}^{\tilde{\ell} + 1}  \mathcal{G}_{\ell}    \Bigg)   L_{1, \tilde{\ell}}  \Bigg[ \chi_{\tilde{k} \geq 2} \Bigg( \prod_{k=2}^{\tilde{k} - 1}  G_{k, \tilde{\ell}} \Bigg) + \chi_{\tilde{k}  = 1} \Bigg( \prod_{k=2}^{W}  G_{k, \tilde{\ell}} \Bigg)   \Bigg] \\
    A_{\tilde{k}, \tilde{\ell}}^+ &=  \Bigg( \prod_{\ell=L}^{\tilde{\ell} + 1}  \mathcal{G}_{\ell}    \Bigg)   L_{1, \tilde{\ell}}  \Bigg[ \chi_{\tilde{k} \geq 2} \Bigg( \prod_{k=2}^{\tilde{k}}  G_{k, \tilde{\ell}} \Bigg) + \chi_{\tilde{k}  = 1} \Bigg( \prod_{k=2}^{W}  G_{k, \tilde{\ell}} \Bigg) R_{1, \tilde{\ell}}   \Bigg] \\
    B_{\tilde{k}, \tilde{\ell}} &= \Bigg[  \chi_{\tilde{k}  \geq 2} \Bigg( \prod_{k=\tilde{k} + 1}^{W}  G_{k, \tilde{\ell}} \Bigg)  R_{1, \tilde{\ell}}  +  \chi_{\tilde{k}  = 1} \mathds{1} \Bigg]  \Bigg( \prod_{\ell=\tilde{\ell} - 1}^{1} \mathcal{G}_{\ell} \Bigg) \\
    T_{\tilde{k}, \tilde{\ell}} &= \chi_{\tilde{k}  \geq 2} \mathrm{SWAP}_{\tilde{k}, \tilde{\ell}} + \chi_{\tilde{k}  = 1} \mathds{1}   \quad 
\end{align*}
and arrive at the expression (we suppress the reference to the current parameterization $s$)
\begin{align*}
    & \rho_{L}^{(x)}(s+ \epsilon) = \rho_{L}^{(x)} \\
    &-i \epsilon \sum_{\tilde{\ell}=1}^{L}  \sum_{\tilde{k}=1}^W \Tr_{\neq L} \Bigg( A_{\tilde{k}, \tilde{\ell}}  T_{\tilde{k}, \tilde{\ell}} \Bigg[ \frac{\sqrt{\delta t}}{2} \tilde{V}_{\tilde{k}, \tilde{\ell}} , e^{-i \sqrt{\delta t} V_{\tilde{k}, \tilde{\ell}}} e^{-i \delta t H_{\tilde{k}, \tilde{\ell}}}  B_{\tilde{k}, \tilde{\ell}}  \Bigg[ \rho_0^{(x)} \otimes \bigotimes_{i = 1}^{L} \ket{\mathbf{0}}\bra{\mathbf{0}} \Bigg] B_{\tilde{k}, \tilde{\ell}}^{ \dagger}  e^{i \delta t H_{\tilde{k}, \tilde{\ell}}} e^{i \sqrt{\delta t} V_{\tilde{k}, \tilde{\ell}}}  \Bigg] T_{\tilde{k}, \tilde{\ell}} A_{\tilde{k}, \tilde{\ell}}^{ \dagger} \\
    &+ A_{\tilde{k}, \tilde{\ell}}  T_{\tilde{k}, \tilde{\ell}} e^{-i \sqrt{\delta t} V_{\tilde{k},  \tilde{\ell}}} \Bigg[ \frac{\sqrt{\delta t}}{2} \tilde{V}_{\tilde{k},\tilde{\ell}} ,  e^{-i \delta t H_{\tilde{k}, \tilde{\ell}}}  B_{\tilde{k}, \tilde{\ell}} \Bigg[ \rho_0^{(x)} \otimes \bigotimes_{i = 1}^{L} \ket{\mathbf{0}}\bra{\mathbf{0}} \Bigg] B_{\tilde{k}, \tilde{\ell}}^{ \dagger}  e^{i \delta t H_{\tilde{k}, \tilde{\ell}}}   \Bigg] e^{i \sqrt{\delta t} V_{\tilde{k},  \tilde{\ell}}} T_{\tilde{k}, \tilde{\ell}} A_{\tilde{k}, \tilde{\ell}}^{ \dagger} \\
    &+ A_{\tilde{k}, \tilde{\ell}}^+    \Bigg[ \delta t \tilde{H}_{\tilde{k},\tilde{\ell}} ,  B_{\tilde{k}, \tilde{\ell}} \Bigg[ \rho_0^{(x)} \otimes \bigotimes_{i = 1}^{L} \ket{\mathbf{0}}\bra{\mathbf{0}} \Bigg]  B_{\tilde{k}, \tilde{\ell}}^{ \dagger}     \Bigg]   \left( A_{\tilde{k}, \tilde{\ell}}^{+ } \right)^\dagger   \Bigg)  +  \mathcal{O}(\epsilon^2) \\
    & =: \rho_{L}^{(x)}(s)  -i \epsilon \Lambda_L^{(x)}(s) +  \mathcal{O}(\epsilon^2) 
\end{align*}
for the updated output states. Here we take the conventions that 
\begin{align*}
     &\prod_{k=2}^{1}  G_{k, \tilde{\ell}}  = \mathds{1}  =\prod_{k=W + 1}^{W}  G_{k, \tilde{\ell}}    \quad , \qquad  \prod_{\ell=L}^{L + 1} \mathcal{G}_{\ell} =  \mathds{1} =  \prod_{\ell=0}^{1} \mathcal{G}_{\ell} \quad .
\end{align*}
The change of the loss to first order in $\epsilon$ then computes as  
\begin{align}\label{losschange}
    \frac{dL(s)}{ds} &=    -\frac{2i}{P} \sum_{x=1}^{P}   \tilde{C}^{(x)}(s)  \Tr_{L} \left( \left( \frac{1}{2W} \sum_{k=1}^W \sigma_k^z - m_{z,\mathrm{out}}^{(x)} \right) \Lambda_L^{(x)}(s) \right) 
\end{align}
with $\Tr_{L}$ the trace over layer $L$ and the magnetization difference for pair $x$ 
\begin{equation*}
    \tilde{C}^{(x)}(s) = \Tr_{L} \left( \left( \frac{1}{2W} \sum_{k=1}^W \sigma_k^z - m_{z,\mathrm{out}}^{(x)} \right) \rho_{L}^{(x)}(s) \right).
\end{equation*}
As now the trace is over all parts, we may use cyclicity to rearrange the second factor in the sum in a more convenient form
\begin{align*}
& \Tr_{L} \bigg( \left( \frac{1}{2W} \sum_{k=1}^W \sigma_k^z - m_{z,\mathrm{out}}^{(x)} \right) \Lambda_L^{(x)}(s) \bigg) \\
=& \sum_{\tilde{\ell},\tilde{k}=1}^{L,W}  \Tr \Bigg( \Bigg( \Bigg[ e^{-i \sqrt{\delta t} V_{\tilde{k}, \tilde{\ell}}} e^{-i \delta t H_{\tilde{k}, \tilde{\ell}}}  B_{\tilde{k}, \tilde{\ell}}  \Bigg[ \rho_0^{(x)} \otimes \bigotimes_{i = 1}^{L} \ket{\mathbf{0}}\bra{\mathbf{0}} \Bigg] B_{\tilde{k}, \tilde{\ell}}^{ \dagger}  e^{i \delta t H_{\tilde{k}, \tilde{\ell}}} e^{i \sqrt{\delta t} V_{\tilde{k}, \tilde{\ell}}} , T_{\tilde{k}, \tilde{\ell}} A_{\tilde{k}, \tilde{\ell}}^{ \dagger} \left( \frac{1}{2W} \sum_{k=1}^W \sigma_k^z - m_{z,\mathrm{out}}^{(x)} \right)  A_{\tilde{k}, \tilde{\ell}}  T_{\tilde{k}, \tilde{\ell}} \Bigg]  \\
    &+ \Bigg[ e^{-i \delta t H_{\tilde{k}, \tilde{\ell}}}  B_{\tilde{k}, \tilde{\ell}} \Bigg[ \rho_0^{(x)} \otimes \bigotimes_{i = 1}^{L} \ket{\mathbf{0}}\bra{\mathbf{0}} \Bigg] B_{\tilde{k}, \tilde{\ell}}^{ \dagger}  e^{i \delta t H_{\tilde{k}, \tilde{\ell}}}  , e^{i \sqrt{\delta t} V_{\tilde{k},  \tilde{\ell}}} T_{\tilde{k}, \tilde{\ell}} A_{\tilde{k}, \tilde{\ell}}^{ \dagger}  \left( \frac{1}{2W} \sum_{k=1}^W \sigma_k^z - m_{z,\mathrm{out}}^{(x)} \right) A_{\tilde{k}, \tilde{\ell}}  T_{\tilde{k}, \tilde{\ell}} e^{-i \sqrt{\delta t} V_{\tilde{k},  \tilde{\ell}}} \Bigg] \Bigg) \frac{\sqrt{\delta t}}{2} \tilde{V}_{\tilde{k},\tilde{\ell}}   \\
    &+ \Bigg[  B_{\tilde{k}, \tilde{\ell}} \Bigg[ \rho_0^{(x)} \otimes \bigotimes_{i = 1}^{L} \ket{\mathbf{0}}\bra{\mathbf{0}} \Bigg]  B_{\tilde{k}, \tilde{\ell}}^{ \dagger} , \left( A_{\tilde{k} , \tilde{\ell}}^{+ } \right)^\dagger \left( \frac{1}{2W} \sum_{k=1}^W \sigma_k^z - m_{z,\mathrm{out}}^{(x)} \right) A_{\tilde{k} , \tilde{\ell}}^{+}    \Bigg] \delta t \tilde{H}_{\tilde{k},\tilde{\ell}}          \Bigg) \\
    =& \sum_{\tilde{\ell},\tilde{k}=1}^{L,W} \sum_{n=1}^3  \Tr \left( \Bigg[ Y_{\tilde{k}, \tilde{\ell}}^{[n]}  B_{\tilde{k}, \tilde{\ell}} \Bigg[ \rho_0^{(x)} \otimes \bigotimes_{i = 1}^{L} \ket{\mathbf{0}}\bra{\mathbf{0}} \Bigg] B_{\tilde{k}, \tilde{\ell}}^{ \dagger}  \left( Y_{\tilde{k}, \tilde{\ell}}^{ [n]} \right)^{\dagger}  , \left( X_{\tilde{k}, \tilde{\ell}}^{[n]}\right)^{ \dagger} A_{\tilde{k}, \tilde{\ell}}^{ \dagger}  \hat{E}_z^{(x)} A_{\tilde{k}, \tilde{\ell}}  X_{\tilde{k}, \tilde{\ell}}^{[n]} \Bigg] \tilde{D}_{\tilde{k}, \tilde{\ell}}^{[n]} \right) \\
    =& \sum_{\tilde{\ell},\tilde{k}=1}^{L,W} \sum_{n=1}^3  \Tr \left( \Bigg[\underbrace{ Y_{\tilde{k}, \tilde{\ell}}^{[n]}  \tilde{B}_{\tilde{k}, \tilde{\ell}} \mathcal{G}_{ \tilde{\ell} - 1}^{\uparrow} }_{\mathrm{support \ up \ to \ \tilde{\ell} }} \Bigg[ \rho_0^{(x)} \otimes \bigotimes_{i = 1}^{L} \ket{\mathbf{0}}\bra{\mathbf{0}} \Bigg] \underbrace{ \mathcal{G}_{ \tilde{\ell} - 1}^{\uparrow \ \dagger} \tilde{B}_{\tilde{k}, \tilde{\ell}}^{ \dagger}  \left( Y_{\tilde{k}, \tilde{\ell}}^{[n]} \right)^{ \dagger} }_{\mathrm{support \ up \ to \  \tilde{\ell} }} , \underbrace{ \left( X_{\tilde{k}, \tilde{\ell}}^{[n]} \right)^{ \dagger} \tilde{A}_{\tilde{k}, \tilde{\ell}}^{ \dagger} \mathcal{G}_{ \tilde{\ell}}^{\downarrow \ \dagger} \hat{E}_z^{(x)} \mathcal{G}_{ \tilde{\ell}}^{\downarrow}  \tilde{A}_{\tilde{k}, \tilde{\ell}}  X_{\tilde{k}, \tilde{\ell}}^{[n]} }_{\mathrm{support \ down \ to \   \tilde{\ell} - 1 }} \Bigg] \tilde{D}_{\tilde{k}, \tilde{\ell}}^{[n]} \right) .
\end{align*}
Here we have further defined 
\begin{align*}
    &\hat{E}_z^{(x)} =  \frac{1}{2W} \sum_{k=1}^W \sigma_k^z - m_{z,\mathrm{out}}^{(x)} ,\\ 
    &X_{\tilde{k}, \tilde{\ell}}^{[1]} = T_{\tilde{k}, \tilde{\ell}} , \quad X_{\tilde{k}, \tilde{\ell}}^{[2]} =  T_{\tilde{k}, \tilde{\ell}} e^{-i \sqrt{\delta t} V_{\tilde{k},  \tilde{\ell}}} , \quad X_{\tilde{k}, \tilde{\ell}}^{[3]} =  \chi_{\tilde{k} \geq 2} G_{\tilde{k}, \tilde{\ell}} + \chi_{\tilde{k} = 1} R_{1, \tilde{\ell}} , \\
    &Y_{\tilde{k}, \tilde{\ell}}^{[1]} = e^{-i \sqrt{\delta t} V_{\tilde{k}, \tilde{\ell}}} e^{-i \delta t H_{\tilde{k}, \tilde{\ell}}} , \quad Y_{\tilde{k}, \tilde{\ell}}^{[2]} = e^{-i \delta t H_{\tilde{k}, \tilde{\ell}}} , \quad Y_{\tilde{k}, \tilde{\ell}}^{[3]} =  \mathds{1},\\ 
    &  \tilde{D}_{\tilde{k}, \tilde{\ell}}^{[1]}  = \tilde{D}_{\tilde{k}, \tilde{\ell}}^{[2]} = \frac{\sqrt{\delta t}}{2} \tilde{V}_{\tilde{k},\tilde{\ell}} , \qquad D_{\tilde{k}, \tilde{\ell}}^{[3]}  = \delta t \tilde{H}_{\tilde{k},\tilde{\ell}},
\end{align*}
and
\begin{align*}
    \mathcal{G}_{ \tilde{\ell}}^{\downarrow}  =& \Bigg( \prod_{\ell=L}^{\tilde{\ell} + 1}  \mathcal{G}_{\ell}    \Bigg)  , \qquad \mathcal{G}_{ \tilde{\ell} - 1}^{\uparrow}  =  \Bigg( \prod_{\ell=\tilde{\ell} - 1}^{1}  \mathcal{G}_{\ell}    \Bigg) , \\
    \tilde{A}_{\tilde{k}, \tilde{\ell}} =& L_{1, \tilde{\ell}}  \Bigg[ \chi_{\tilde{k} \geq 2} \Bigg( \prod_{k=2}^{\tilde{k} - 1}  G_{k, \tilde{\ell}} \Bigg) + \chi_{\tilde{k}  = 1} \Bigg( \prod_{k=2}^{W}  G_{k, \tilde{\ell}} \Bigg)  \Bigg] , \qquad \tilde{B}_{\tilde{k}, \tilde{\ell}} =  \chi_{\tilde{k}  \geq 2} \Bigg( \prod_{k=\tilde{k} + 1}^{W}  G_{k, \tilde{\ell}} \Bigg)   R_{1, \tilde{\ell}}  +  \chi_{\tilde{k}  = 1} \mathds{1} . 
\end{align*}
Under the trace, due to cyclicity and the fact that operators with different support commute, the vacuum states on layers larger than or equal to $\tilde{\ell} + 1$ can just be pulled to the right-hand side of the commutator. Moreover, the left-hand side of the commutator is supported only up to layer $\tilde{\ell}$ and the right-hand side down to layer $\tilde{\ell} - 1$ such that we can separate the trace and obtain the expression 
\begin{align*}
    &\Tr_{L} \bigg( \bigg( \frac{1}{2W} \sum_{k=1}^W \sigma_k^z - m_{z,\mathrm{out}}^{(x)} \bigg) \Lambda_L^{(x)}(s) \bigg) \\
=& \sum_{\tilde{\ell},\tilde{k}=1}^{L,W} \sum_{n=1}^3  \Tr \Bigg( \Bigg[ Y_{\tilde{k}, \tilde{\ell}}^{[n]}  \tilde{B}_{\tilde{k}, \tilde{\ell}} \Tr_{ \leq \tilde{\ell} - 2} \Bigg( \mathcal{G}_{ \tilde{\ell} - 1}^{\uparrow}  \Bigg[ \rho_0^{(x)} \otimes \bigotimes_{i = 1}^{\tilde{\ell}} \ket{\mathbf{0}}\bra{\mathbf{0}} \Bigg]  \mathcal{G}_{ \tilde{\ell} - 1}^{\uparrow \ \dagger} \Bigg) \tilde{B}_{\tilde{k}, \tilde{\ell}}^{ \dagger}  \left( Y_{\tilde{k}, \tilde{\ell}}^{[n]} \right)^{ \dagger} , \\
&\left( X_{\tilde{k}, \tilde{\ell}}^{[n]} \right)^{ \dagger} \tilde{A}_{\tilde{k}, \tilde{\ell}}^{ \dagger} \Tr_{\geq  \tilde{\ell} + 1} \Bigg( \Bigg[  \bigotimes_{i = \tilde{\ell}-1}^{\tilde{\ell}}   \mathds{1}    \otimes \bigotimes_{i=\tilde{\ell} + 1}^{L} \ket{\mathbf{0}}\bra{\mathbf{0}} \Bigg] \mathcal{G}_{ \tilde{\ell}}^{\downarrow \ \dagger} \hat{E}_z^{(x)} \mathcal{G}_{ \tilde{\ell}}^{\downarrow} \Bigg) \tilde{A}_{\tilde{k}, \tilde{\ell}}  X_{\tilde{k}, \tilde{\ell}}^{[n]}  \Bigg] \tilde{D}_{\tilde{k}, \tilde{\ell}}^{[n]} \Bigg) \\
=& \sum_{\tilde{\ell},\tilde{k}=1}^{L,W} \sum_{n=1}^3  \Tr \Bigg( \Bigg[ Y_{\tilde{k}, \tilde{\ell}}^{[n]}  \tilde{B}_{\tilde{k}, \tilde{\ell}} \Bigg( \rho_{\tilde{\ell}-1}^{(x)} \otimes \outerproduct{\mathbf{0
     }}{\mathbf{0}} \Bigg) \tilde{B}_{\tilde{k}, \tilde{\ell}}^{ \dagger}  \left( Y_{\tilde{k}, \tilde{\ell}}^{[n]} \right)^{ \dagger} , \left( X_{\tilde{k}, \tilde{\ell}}^{[n]} \right)^{ \dagger} \tilde{A}_{\tilde{k}, \tilde{\ell}}^{ \dagger} \Bigg( \mathds{1} \otimes \sigma_{\tilde{\ell}}^{(x)} \Bigg) \tilde{A}_{\tilde{k}, \tilde{\ell}}  X_{\tilde{k}, \tilde{\ell}}^{[n]}  \Bigg] \tilde{D}_{\tilde{k}, \tilde{\ell}}^{[n]} \Bigg) .
\end{align*}
The reduced state $\rho_{\tilde{l}-1}^{(x)}$ can be obtained by propagating the initial state labeled by $x$ according to Eq.~\eqref{recurrence_sup}. The reduced state $\sigma_{\tilde{l}}^{(x)}$ can be obtain by 'backpropagating' the operator $\hat{E}_z^{(x)}$ via the recurrence relation
\begin{equation} \label{recurrence_2_sup}
    \sigma_\ell^{(x)}(s) = \Tr_{\ell+1}\left(  \mathds{1} \otimes \ket{\mathbf{0}}\bra{\mathbf{0}}  \mathcal{G}_{\ell}^\dagger(s) \mathds{1} \otimes \sigma_{\ell+1}^{(x)}(s)  \mathcal{G}_{\ell}(s) \right) .
\end{equation}
In the special cases that $\tilde{l}=1 \ (W)$ there cannot be pulled a trace inside the left-hand (right-hand) side of the commutator and the reduced state is just $\rho_0^{(x)}$ ($\hat{E}_z^{(x)}$). Thus, the above expression is an efficient one in terms of the number of layers $L$. With Eq.~\eqref{losschange}, the change of the loss becomes 
\begin{align*}
        \frac{dL}{ds} &= -  \frac{2i}{P} \sum_{x=1}^{P} \sum_{\tilde{\ell}=1}^{L} \sum_{\tilde{k}=1}^W \sum_{n=1}^3  \tilde{C}^{(x)}(s)    \Tr \Bigg( \Bigg[ Y_{\tilde{k}, \tilde{\ell}}^{[n]}  \tilde{B}_{\tilde{k}, \tilde{\ell}} \Bigg( \rho_{\tilde{\ell}-1}^{(x)} \otimes \outerproduct{\mathbf{0
     }}{\mathbf{0}} \Bigg) \tilde{B}_{\tilde{k}, \tilde{\ell}}^{ \dagger}  \left( Y_{\tilde{k}, \tilde{\ell}}^{[n]} \right)^{ \dagger} , \left( X_{\tilde{k}, \tilde{\ell}}^{[n]} \right)^{ \dagger} \tilde{A}_{\tilde{k}, \tilde{\ell}}^{ \dagger} \Bigg( \mathds{1} \otimes \sigma_{\tilde{\ell}}^{(x)} \Bigg) \tilde{A}_{\tilde{k}, \tilde{\ell}}  X_{\tilde{k}, \tilde{\ell}}^{[n]}  \Bigg] \tilde{D}_{\tilde{k}, \tilde{\ell}}^{[n]} \Bigg) \\
    &= -  \frac{2i}{P} \sum_{x=1}^{P}    \sum_{\tilde{\ell}=1}^{L}  \sum_{\tilde{k}=1}^W  \tilde{C}^{(x)}(s) \Tr\Bigg( M_{\tilde{k}, \tilde{\ell}}^{ (x)} \frac{\sqrt{\delta t}}{2} \tilde{V}_{\tilde{k}, \tilde{\ell}} + N_{\tilde{k}, \tilde{\ell}}^{ (x)} \delta t \tilde{H}_{\tilde{k},\tilde{\ell}}  \Bigg)  ,
\end{align*}
with 
\begin{align*}
    M_{\tilde{k}, \tilde{\ell}}^{ (x)} =& \Bigg[ Y_{\tilde{k}, \tilde{\ell}}^{[1]}  \tilde{B}_{\tilde{k}, \tilde{\ell}} \Bigg( \rho_{\tilde{\ell}-1}^{(x)} \otimes \outerproduct{\mathbf{0
     }}{\mathbf{0}} \Bigg) \tilde{B}_{\tilde{k}, \tilde{\ell}}^{ \dagger}  \left( Y_{\tilde{k}, \tilde{\ell}}^{[1]} \right)^{ \dagger} , \left( X_{\tilde{k}, \tilde{\ell}}^{[1]} \right)^{ \dagger} \tilde{A}_{\tilde{k}, \tilde{\ell}}^{ \dagger} \Bigg( \mathds{1} \otimes \sigma_{\tilde{\ell}}^{(x)} \Bigg) \tilde{A}_{\tilde{k}, \tilde{\ell}}  X_{\tilde{k}, \tilde{\ell}}^{[1]}  \Bigg] \\
     &+ \Bigg[ Y_{\tilde{k}, \tilde{\ell}}^{[2]}  \tilde{B}_{\tilde{k}, \tilde{\ell}} \Bigg( \rho_{\tilde{\ell}-1}^{(x)} \otimes \outerproduct{\mathbf{0
     }}{\mathbf{0}} \Bigg) \tilde{B}_{\tilde{k}, \tilde{\ell}}^{ \dagger}  \left( Y_{\tilde{k}, \tilde{\ell}}^{[2]} \right)^{ \dagger} , \left( X_{\tilde{k}, \tilde{\ell}}^{[2]} \right)^{ \dagger} \tilde{A}_{\tilde{k}, \tilde{\ell}}^{ \dagger} \Bigg( \mathds{1} \otimes \sigma_{\tilde{\ell}}^{(x)} \Bigg) \tilde{A}_{\tilde{k}, \tilde{\ell}}  X_{\tilde{k}, \tilde{\ell}}^{[2]}  \Bigg] ,\\
    N_{\tilde{k}, \tilde{\ell}}^{ (x)} =& \Bigg[ Y_{\tilde{k}, \tilde{\ell}}^{[3]}  \tilde{B}_{\tilde{k}, \tilde{\ell}} \Bigg( \rho_{\tilde{\ell}-1}^{(x)} \otimes \outerproduct{\mathbf{0
     }}{\mathbf{0}} \Bigg) \tilde{B}_{\tilde{k}, \tilde{\ell}}^{ \dagger}  \left( Y_{\tilde{k}, \tilde{\ell}}^{[3]} \right)^{ \dagger} , \left( X_{\tilde{k}, \tilde{\ell}}^{[3]} \right)^{ \dagger} \tilde{A}_{\tilde{k}, \tilde{\ell}}^{ \dagger} \Bigg( \mathds{1} \otimes \sigma_{\tilde{\ell}}^{(x)} \Bigg) \tilde{A}_{\tilde{k}, \tilde{\ell}}  X_{\tilde{k}, \tilde{\ell}}^{[3]}  \Bigg] .
\end{align*}
This is a linear function of the update parameters $\tilde{c}^{\vec{\alpha}}= \tilde{x}^{\vec{\alpha}} + i \tilde{y}^{\vec{\alpha}}, \tilde{d}^{\vec{\alpha}}$:
\begin{align*}
        \frac{dL}{ds} =& -  \frac{2i}{P} \sum_{x=1}^{P}    \sum_{\tilde{\ell}=1}^{L}  \sum_{\tilde{k}=1}^W  \tilde{C}^{(x)}(s) \Tr\Bigg( M_{\tilde{k}, \tilde{\ell}}^{ (x)} \frac{\sqrt{\delta t}}{2} \tilde{V}_{\tilde{k}, \tilde{\ell}} + N_{\tilde{k}, \tilde{\ell}}^{ (x)} \delta t \tilde{H}_{\tilde{k},\tilde{\ell}}  \Bigg) \\
        =& -  \frac{2i}{P} \sum_{x=1}^{P}    \sum_{\tilde{\ell}=1}^{L}  \sum_{\tilde{k}=1}^W  \tilde{C}^{(x)}(s)  \Tr\Bigg( M_{\tilde{k}, \tilde{\ell}}^{ (x)} \frac{\sqrt{\delta t}}{2} \sum_{\vec{\alpha}} \Big(  \Big( \tilde{x}_{\tilde{k}}^{\vec{\alpha}}+i\tilde{y}_{\tilde{k}}^{\vec{\alpha}} \Big) \hat{S}^{+,\vec{\alpha}}_{\tilde{k},\tilde{\ell}} + \Big( \tilde{x}_{\tilde{k}}^{\vec{\alpha}}-i\tilde{y}_{\tilde{k}}^{\vec{\alpha}} \Big) \hat{S}^{-,\vec{\alpha}}_{\tilde{k},\tilde{\ell}} \Big) \Bigg) \\
        & -  \frac{2i}{P} \sum_{x=1}^{P}    \sum_{\tilde{\ell}=1}^{L}  \sum_{\tilde{k}=1}^W  \tilde{C}^{(x)}(s) \Tr \Bigg( N_{\tilde{k}, \tilde{\ell}}^{ (x)} \delta t \sum_{\vec{\alpha}} \tilde{d}_{\tilde{k}}^{\vec{\alpha}} \hat{S}^{\vec{\alpha}}_{\tilde{k},\tilde{\ell}}   \Bigg) \\
        =& -   \frac{i \sqrt{\delta t}}{P} \sum_{x=1}^{P}    \sum_{\tilde{\ell}=1}^{L}  \sum_{\tilde{k}=1}^W  \sum_{\vec{\alpha}} \tilde{C}^{(x)}(s)  \Tr\Bigg( M_{\tilde{k}, \tilde{\ell}}^{ (x)}   \Big(   \hat{S}^{+,\vec{\alpha}}_{\tilde{k},\tilde{\ell}} +  \hat{S}^{-,\vec{\alpha}}_{\tilde{k},\tilde{\ell}} \Big)  \Bigg) \Big( \chi_{\tilde{k} \geq 2}  \tilde{x}^{\vec{\alpha}} +  \chi_{\tilde{k} = 1}  \delta^{\alpha_1 0}  \tilde{x}^{\vec{\alpha}}  \Big) \\
        &+  \frac{\sqrt{\delta t}}{P} \sum_{x=1}^{P}    \sum_{\tilde{\ell}=1}^{L}  \sum_{\tilde{k}=1}^W  \sum_{\vec{\alpha}} \tilde{C}^{(x)}(s)  \Tr\Bigg( M_{\tilde{k}, \tilde{\ell}}^{ (x)}   \Big(   \hat{S}^{+,\vec{\alpha}}_{\tilde{k},\tilde{\ell}} - \hat{S}^{-,\vec{\alpha}}_{\tilde{k},\tilde{\ell}} \Big) \Bigg) \Big( \chi_{\tilde{k} \geq 2}  \tilde{y}^{\vec{\alpha}} +  \chi_{\tilde{k} = 1}  \delta^{\alpha_1 0}  \tilde{y}^{\vec{\alpha}}  \Big) \\
        & -  \frac{2 i\delta t}{P} \sum_{x=1}^{P}    \sum_{\tilde{\ell}=1}^{L} \sum_{\tilde{k}=1}^W   \sum_{\vec{\alpha}} \tilde{C}^{(x)}(s) \Tr \Bigg( N_{\tilde{k}, \tilde{\ell}}^{ (x)}  \hat{S}^{\vec{\alpha}}_{\tilde{k},\tilde{\ell}}   \Bigg) \Big( \chi_{\tilde{k} \geq 2}  \tilde{d}^{\vec{\alpha}} +  \chi_{\tilde{k} = 1}  \delta^{\alpha_1 0} \tilde{d}^{\vec{\alpha}}  \Big).
\end{align*}
We arrive at the update parameters by moving into the negative direction of the gradient
\begin{align*}
    \tilde{d}^{\vec{\alpha}} &=   \frac{2 i  \delta t}{P} \sum_{x=1}^{P}    \sum_{\tilde{\ell}=1}^{L} \sum_{\tilde{k}=1}^W  \tilde{C}^{(x)}(s) \Tr \Bigg( N_{\tilde{k}, \tilde{\ell}}^{ (x)}   \hat{S}^{\vec{\alpha}}_{\tilde{k},\tilde{\ell}}   \Bigg) \Bigg[ \chi_{\tilde{k} \geq 2}  + \chi_{\tilde{k} = 1} \delta^{\alpha_1 0}  \Bigg] ,\\
    \tilde{x}^{\vec{\alpha}} &=    \frac{i  \sqrt{\delta t}}{P} \sum_{x=1}^{P}    \sum_{\tilde{\ell}=1}^{L} \sum_{\tilde{k}=1}^W  \tilde{C}^{(x)}(s) \Tr \Bigg( M_{\tilde{k}, \tilde{\ell}}^{ (x)} \Big(   \hat{S}^{+,\vec{\alpha}}_{\tilde{k},\tilde{\ell}} +  \hat{S}^{-,\vec{\alpha}}_{\tilde{k},\tilde{\ell}} \Big) \Bigg) \Bigg[ \chi_{\tilde{k} \geq 2}  + \chi_{\tilde{k} =1} \delta^{\alpha_1 0}   \Bigg] ,\\
    \tilde{y}^{\vec{\alpha}} &=-    \frac{ \sqrt{\delta t}}{P} \sum_{x=1}^{P}    \sum_{\tilde{\ell}=1}^{L} \sum_{\tilde{k}=1}^W  \tilde{C}^{(x)}(s) \Tr \Bigg( M_{\tilde{k}, \tilde{\ell}}^{ (x)} \Big(   \hat{S}^{+,\vec{\alpha}}_{\tilde{k},\tilde{\ell}} - \hat{S}^{-,\vec{\alpha}}_{\tilde{k},\tilde{\ell}} \Big)  \Bigg) \Bigg[ \chi_{\tilde{k} \geq 2}  + \chi_{\tilde{k} = 1} \delta^{\alpha_1 0}   \Bigg] .
\end{align*}
The amount by which we move into this direction is controlled by the global learning rate $\epsilon$.

\section{Tensor network methods}
To summarize, we found ways to efficiently express (in terms of $L$) the input-output relation and the update parameters via the recurrence relations in Eqs.~\eqref{recurrence_sup},\eqref{recurrence_2_sup}. Still, for large widths $W$ it is difficult to compute them classically due to exponential growth of Hilbert space with system size. In the following we present an approximate way based on tensor networks \cite{gillman2021b}. For this we need to distinguish the cases (we omit the label $x$)
\begin{align*}
\mathrm{(1)} & \qquad \rho_\ell = \Tr_{\ell-1}(\mathcal{G}_{\ell} \rho_{\ell-1} \otimes \ket{\mathbf{0}}\bra{\mathbf{0}}  \mathcal{G}_{\ell}^\dagger) = \mathcal{F}^{\ell-1,\ell}\left[ \rho_{\ell-1} \right] \\
\mathrm{(2)} & \qquad   \sigma_\ell = \Tr_{\ell+1}\left(\mathds{1} \otimes \outerproduct{\mathbf{0}}{\mathbf{0}} \mathcal{G}_{\ell}^\dagger \mathds{1} \otimes \sigma_{\ell+1}  \mathcal{G}_{\ell} \right) = \mathcal{B}^{\ell+1,\ell} \left[ \sigma_{\ell+1} \right] \\
\mathrm{(3)} & \qquad  \Tr \Bigg( N_{\tilde{k}, \tilde{\ell}}  \hat{S}^{\vec{\alpha}}_{\tilde{k},\tilde{\ell}}   \Bigg) \\
\mathrm{(4)} & \qquad \Tr \Bigg( M_{\tilde{k}, \tilde{\ell}} \Big(   \hat{S}^{+,\vec{\alpha}}_{\tilde{k},\tilde{\ell}} \pm  \hat{S}^{-,\vec{\alpha}}_{\tilde{k},\tilde{\ell}} \Big) \Bigg)
\end{align*} 
The first is just the recurrence relation \eqref{recurrence_sup} which propagates the input. We call the corresponding channel $\mathcal{F}^{\ell-1,\ell}$ the forward channel. The second case is the recurrence relation \eqref{recurrence_2_sup} needed in the update, where the error operator $\hat{E}_z$ is 'backpropagated'. We call the channel $\mathcal{B}^{\ell+1,\ell} $ the backward channel. The last two cases concern the last step in the calculation of the update parameters where a number is computed based on the forward and backward propagated states. All the cases may equivalently be expressed via vector states and operators acting on them. Indeed, note that we can use the isomorphism 
\begin{align}\label{dsiso}
O = \sum_{\vec{l}, \vec{r}} O^{\vec{l}\vec{r}} \outerproduct{\vec{l}}{\vec{r}} \longrightarrow \ket{O} = \sum_{\vec{l}, \vec{r}} O^{\vec{l}\vec{r}} \ket{\vec{l}} \otimes \ket{\vec{r}}
\end{align}
to express a generic operator $O$ as a vector $\ket{O}$ in a 'doubled-space'. Here $\ket{\vec{l}}, \ket{\vec{r}}$ denote computational basis states on a full layer and the sum over all of these.
\subsection{Cases (1) and (2)}
Using \eqref{dsiso} to vectorize cases (1) and (2), we arrive at the identities 
\begin{align*}
    \mathrm{(1)'} & \qquad \ket{\rho_\ell} = \bra{\mathds{1}} \left( \mathcal{G}_{\ell} \otimes \mathcal{G}_\ell^* \right) \left( \mathds{1} \otimes \ket{\mathbf{0}} \otimes \mathds{1} \otimes \ket{\mathbf{0}} \right) \ket{\rho_{\ell-1}} = F^{\ell-1,\ell} \ket{\rho_{\ell-1}} \\
    \mathrm{(2)'} & \qquad \ket{\sigma_\ell} = \left( \mathds{1} \otimes \bra{\mathbf{0}} \otimes \mathds{1} \otimes \bra{\mathbf{0}} \right)   \left( \mathcal{G}_{\ell+1}^\dagger \otimes \mathcal{G}_{\ell+1}^T \right) \ket{\mathds{1}} \ket{\sigma_{\ell+1}}  = B^{\ell,\ell+1} \ket{\sigma_{\ell+1}} ,
\end{align*}
\begin{figure*}[t]
\centering
    \includegraphics[width=0.9\textwidth]{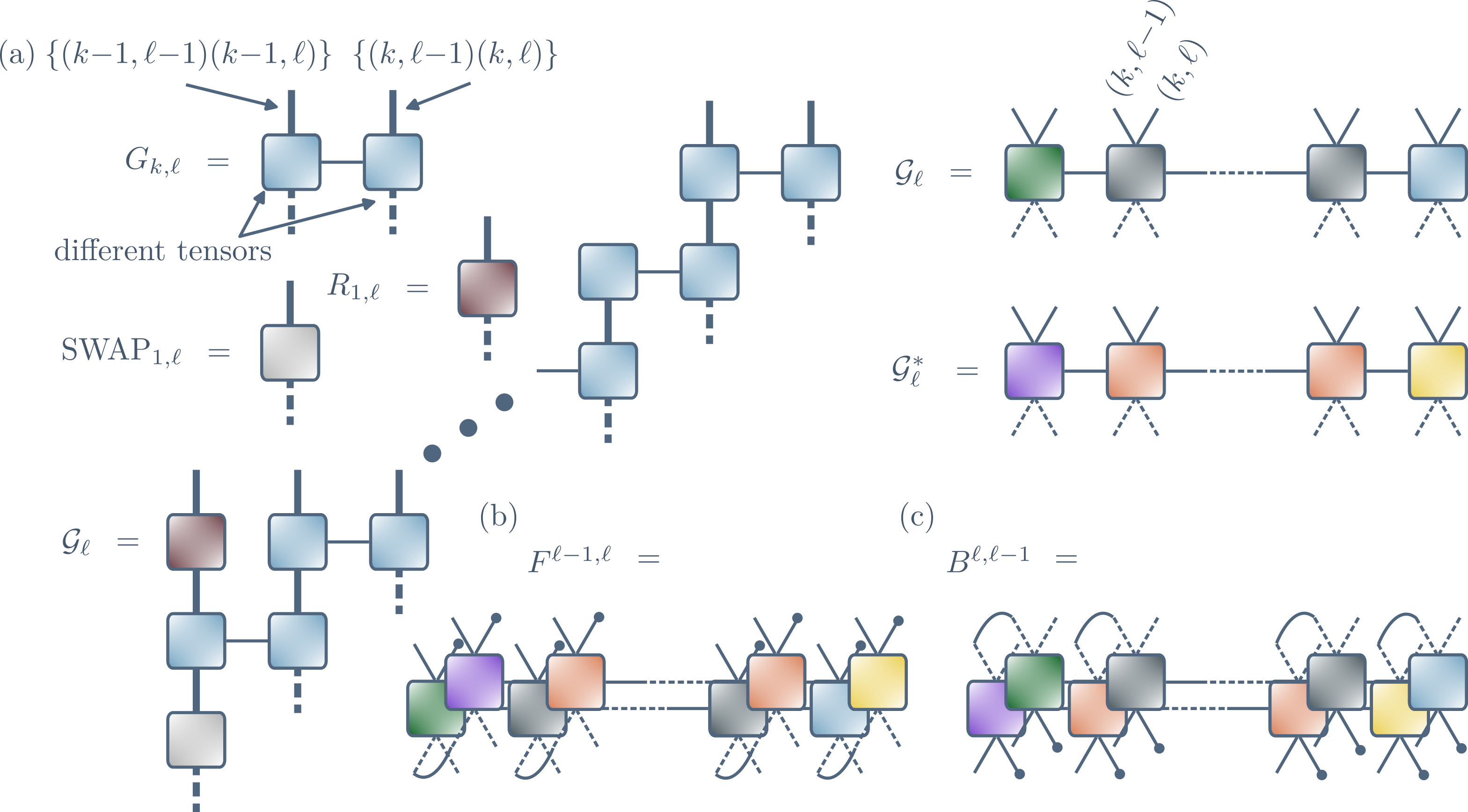}
    \caption{\textbf{MPOs for the channels.} (a) The translation invariant gates $G_{k,\ell}$ can be written as MPOs employing, for instance, QR-decomposition. Here we joined the physical indices associated to site $k$ on layer $\ell-1$ and site $k$ on layer $\ell$. Note that the gate only acts on sites $k-1,k$ of two adjacent layers. Downward pointing legs correspond to outgoing indices and upward pointing ones to ingoing indices. Connected legs represent virtual indices (with bond dimension $\chi$) that are contracted. Due to our open boundary conditions, the gates $\mathrm{SWAP}_{1,\ell}, R_{1,\ell}$ are already in MPO form. Multiplying them, we get the MPO for $\mathcal{G}_\ell$ and $\mathcal{G}_\ell^*$ (conjugating the matrices); the central matrices (grey and orange, respectively) are translation invariant. (b) Based on $\mathcal{G}_\ell$ and $\mathcal{G}_\ell^*$ we get the MPO $F^{\ell-1,\ell}$ (acting on the doubled-space) by evaluating the ingoing indices for $(k,\ell),[k,\ell]$ [see (1)'] for the vacuum. The outgoing indices $(k,\ell-1),[k,\ell-1]$ are contracted. (c) For $B^{\ell,\ell-1}$ in contrast, the MPO is transposed and the order of the physical indices is reversed.}
    \label{Fig1_sup}
\end{figure*}
which are operator-vector state products with 
\begin{align*}
    \ket{\mathds{1}} = \sum_{\vec{\epsilon}} \ket{\vec{\epsilon}} \otimes \mathds{1} \otimes \ket{\vec{\epsilon}}  \otimes \mathds{1} .
\end{align*}
Here $\ket{\vec{\epsilon}} = \ket{\epsilon_1,\epsilon_2,\ldots,\epsilon_W}$ is a computational basis state on one layer and the sum is over all $\epsilon_i =0,1, \ i=1,\ldots,W$. We can find a way to reduce the number of degrees of freedom involved in these operator-state products by expressing $F^{\ell-1,\ell}$ and $B^{\ell+1,\ell}$ as MPOs [see Fig.~\ref{Fig1_sup}(b),(c)]. This can be done based on the MPOs for $\mathcal{G}_\ell$ and $\mathcal{G}_\ell^*$ which, in turn, can be obtained from the MPOs of the local gates as illustrated in Fig.~\ref{Fig1_sup}(a). Note that we can freely choose the ordering of the sites in $F^{\ell-1,\ell}$ and $B^{\ell+1,\ell}$ (as long as it is consistent for all operators and states involved) and we order them as 
\begin{equation*}
(1,\ell-1),[1,\ell-1],(1,\ell),[1,\ell],\ldots, (k,\ell-1),[k,\ell-1],(k,\ell),[k,\ell],\ldots,(W,\ell-1),[W,\ell-1],(W,\ell),[W,\ell]
\end{equation*}
where the notation $(k,\ell)$ refers to site $k$ on layer $\ell$ and $[k,\ell]$ to the same in the space introduced by the doubled-space isomorphism. From this it follows that the reduced states $\rho_\ell = \sum_{\vec{l},\vec{r}} C^{\vec{l}\vec{r}} \outerproduct{\vec{l}}{\vec{r}}$ are written in doubled space as $\ket{\rho_\ell }= \sum_{\vec{l},\vec{r}} C^{\vec{l}\vec{r}} \ket{l_1r_1l_2r_2\ldots l_W r_W}$ (and similarly $\sigma_\ell$). We can join the indices $l_k,r_k$ and name the elements of the product basis according to $00\rightarrow 0, 01 \rightarrow 1, 10 \rightarrow 2, 11 \rightarrow 3$. \\
Generally, the MPOs for $F^{\ell-1,\ell}$ and $B^{\ell+1,\ell}$ are repeatedly applied to the vectorized reduced states and we can approximate this using standard MPS compression algorithms \cite{paeckel2019}, provided the state is an MPS. Note that if the input state is an MPS then the result of an MPO-MPS application is an MPS as well. \\
For forward propagation, the input states we are employing are translation invariant product states that are characterized by a certain value for the magnetization $m_z$ (see main text). Their (translation invariant) tensors of the MPS are given by
\begin{align*}
    Q^{0} &= \cos (\theta / 2)^2 \\
    Q^{1} &= \cos (\theta / 2) e^{-i \phi} \sin  (\theta / 2) \\
    Q^{2} &= \cos (\theta / 2) e^{i \phi} \sin  (\theta / 2)\\
    Q^{3} &= \sin  (\theta / 2)^2 
\end{align*}
with $\theta = \arccos (2 m_z)$ and $m^x = \frac{1}{2} \sin \theta \cos \phi$. For backward propagation, the input states are vectorized error operators
\begin{equation*}
    \hat{E}_z = \frac{1}{2W} \sum_{k=1}^W \sigma_k^z - m_{z,\mathrm{out}} \mathds{1} .
\end{equation*}
With $W_z^\tau = (\delta^{\tau 0} - \delta^{\tau 3}) /(2W)$ and $U^\tau = \delta^{\tau 0} + \delta^{\tau 3}$ the tensors in their MPS assume the forms
\begin{align*}
    R^{\tau_1} &= \begin{pmatrix} W_z^{\tau_1} & U^{\tau_1} \end{pmatrix}\\
    R^{\tau_2} &= \begin{pmatrix} U^{\tau_2} & 0 & 0 \\ 0 & W_z^{\tau_2} & U^{\tau_2}  \end{pmatrix}\\
    R^{\tau_3} &= \begin{pmatrix} U^{\tau_3} & 0 & 0 & 0 \\  0 & U^{\tau_3} & 0 & 0 \\ 0 & 0 & W_z^{\tau_3} & U^{\tau_3}  \end{pmatrix}  \\
    \vdots & \hspace{100pt} \vdots \\
    R^{\tau_{W-1}} &= \begin{pmatrix} U^{\tau_{W-1}} & \cdots & 0 & 0 & 0 & 0 \\ \vdots & \ddots & \vdots & \vdots & \vdots & \vdots \\ 0 & \cdots &  U^{\tau_{W-1}} & 0 & 0 & 0 \\  0 & \cdots & 0 &  U^{\tau_{W-1}} & 0 & 0 \\ 0 & \cdots & 0& 0 & W_z^{\tau_{W-1}} &  U^{\tau_{W-1}} \end{pmatrix} \\
    R^{\tau_{W}} &= \begin{pmatrix} U^{\tau_{W}} \\ \vdots \\ U^{\tau_{W}}  \\ W_z^{\tau_W} - m_z U^{\tau_{W}} \end{pmatrix} .
\end{align*}
Moreover, we need to be able to compute magnetizations. Also the expression for $m_z$ can be translated into a doubled-space. In fact, for the state $\rho = \sum_{\vec{l},\vec{r}} \rho^{\vec{l}\vec{r}} \outerproduct{\vec{l}}{\vec{r}} \rightarrow \ket{\rho} = \sum_{\vec{l},\vec{r}} \rho^{\vec{l}\vec{r}} \ket{\vec{l}} \otimes \ket{\vec{r}}$, it holds
\begin{align*}
    m_z = \Tr \left( \left( \frac{1}{2W} \sum_{k=1}^W \sigma_k^z \right) \rho \right) = \frac{1}{2W} \sum_{k=1}^W  \bra{\mathds{1}} \left( \sigma_k^z \otimes \mathds{1} \right) \ket{\rho}
\end{align*}
where $\ket{\mathds{1}}=\sum_{\vec{\epsilon}} \ket{\vec{\epsilon}} \otimes \ket{\vec{\epsilon}}$ and $ \sigma_k^z \otimes \mathds{1} $ can be straightforwardly written as MPS and MPO, respectively, such that this amounts to an MPO-MPS application and an overlap calculation.
\subsection{Cases (3) and (4)}
In (3) and (4) the reduced states are used to calculate the update parameters. We also have to formulate these identities in terms of MPSs and MPOs in doubled-space. For this lets write out the different contributions. We start with (3)
\begin{align*}
    \Tr \Bigg( N_{\tilde{k}, \tilde{ \ell}}   \hat{S}^{\vec{\alpha}}_{\tilde{k},\tilde{ \ell}}   \Bigg) =& \Tr \Bigg( \tilde{A}_{\tilde{k}, \tilde{\ell}}  X_{\tilde{k}, \tilde{\ell}}^{[3]}  \hat{S}^{\vec{\alpha}}_{\tilde{k},\tilde{ \ell}}    Y_{\tilde{k}, \tilde{\ell}}^{[3]}  \tilde{B}_{\tilde{k}, \tilde{\ell}} \Bigg( \rho_{\tilde{\ell}-1} \otimes \outerproduct{\mathbf{0
     }}{\mathbf{0}} \Bigg) \tilde{B}_{\tilde{k}, \tilde{\ell}}^{ \dagger}   \left( X_{\tilde{k}, \tilde{\ell}}^{[3]} Y_{\tilde{k}, \tilde{\ell}}^{[3]}  \right)^{ \dagger} \tilde{A}_{\tilde{k}, \tilde{\ell}}^{ \dagger} \Bigg( \mathds{1} \otimes \sigma_{\tilde{\ell}} \Bigg) \Bigg) \\
     &-\Tr \Bigg( \Bigg( \mathds{1} \otimes \sigma_{\tilde{\ell}} \Bigg) \tilde{A}_{\tilde{k}, \tilde{\ell}}  X_{\tilde{k}, \tilde{\ell}}^{[3]}  Y_{\tilde{k}, \tilde{\ell}}^{[3]}  \tilde{B}_{\tilde{k}, \tilde{\ell}} \Bigg( \rho_{\tilde{\ell}-1} \otimes \outerproduct{\mathbf{0
     }}{\mathbf{0}} \Bigg) \tilde{B}_{\tilde{k}, \tilde{\ell}}^{ \dagger}  \left( Y_{\tilde{k}, \tilde{\ell}}^{[3]} \right)^{ \dagger}  \hat{S}^{\vec{\alpha}}_{\tilde{k},\tilde{ \ell}}  \left( X_{\tilde{k}, \tilde{\ell}}^{[3]} \right)^{ \dagger} \tilde{A}_{\tilde{k}, \tilde{\ell}}^{ \dagger}  \Bigg)\\
    &=   u_{\tilde{k}, \tilde{ \ell}}^{\vec{\alpha}} - \left( u_{\tilde{k}, \tilde{ \ell}}^{\vec{\alpha}} \right)^* = 2 i \Im \Big(u_{\tilde{k}, \tilde{ \ell}}^{\vec{\alpha}} \Big).
\end{align*}
We used that $\Tr(X^\dagger) = \Tr(X)^*$ and that $\sigma_{\tilde{ \ell}}^{\ \dagger} = \sigma_{\tilde{ \ell}}$ (backward channel is positive) and $\rho_{\tilde{ \ell} - 1}^{ \ \dagger} = \rho_{\tilde{ \ell} - 1}$. To summarize, with 
\begin{align*}
    \mathcal{G}_{\tilde{\ell}} =  \tilde{A}_{\tilde{k}, \tilde{\ell}}  X_{\tilde{k}, \tilde{\ell}}^{[3]} Y_{\tilde{k}, \tilde{\ell}}^{[3]}    \tilde{B}_{\tilde{k}, \tilde{\ell}}
\end{align*}
we get
\begin{align*}
    u_{\tilde{k}, \tilde{ \ell}}^{\vec{\alpha}}  =& \Tr \Bigg( \tilde{A}_{\tilde{k}, \tilde{\ell}}  X_{\tilde{k}, \tilde{\ell}}^{[3]}  \hat{S}^{\vec{\alpha}}_{\tilde{k},\tilde{ \ell}}    Y_{\tilde{k}, \tilde{\ell}}^{[3]}  \tilde{B}_{\tilde{k}, \tilde{\ell}} \Bigg( \rho_{\tilde{\ell}-1} \otimes \outerproduct{\mathbf{0
     }}{\mathbf{0}} \Bigg) \mathcal{G}_{\tilde{\ell}}^\dagger \Bigg( \mathds{1} \otimes \sigma_{\tilde{\ell}} \Bigg) \Bigg)   .
\end{align*}
On the other hand, for (4) and with
\begin{align*}
X_{\tilde{k}, \tilde{\ell}}^{[1]}  Y_{\tilde{k}, \tilde{\ell}}^{[1]}&=T_{\tilde{k}, \tilde{\ell}}e^{-i \sqrt{\delta t} V_{\tilde{k}, \tilde{\ell}}} e^{-i \delta t H_{\tilde{k}, \tilde{\ell}}}=X_{\tilde{k}, \tilde{\ell}}^{[2]}  Y_{\tilde{k}, \tilde{\ell}}^{[2]} \quad ,\\
\mathcal{G}_{\tilde{\ell}} =& \tilde{A}_{\tilde{k}, \tilde{\ell}}  X_{\tilde{k}, \tilde{\ell}}^{[1]}  Y_{\tilde{k}, \tilde{\ell}}^{[1]} \tilde{B}_{\tilde{k}, \tilde{\ell}} ,
\end{align*}
we have 
\begin{align*}
    &\Tr \Bigg( M_{\tilde{k}, \tilde{ \ell}} \Big(   \hat{S}^{+,\vec{\alpha}}_{\tilde{k},\tilde{ \ell}} \pm  \hat{S}^{-,\vec{\alpha}}_{\tilde{k},\tilde{ \ell}} \Big) \Bigg) \\ 
    =& \Tr \Bigg( \tilde{A}_{\tilde{k}, \tilde{\ell}}  X_{\tilde{k}, \tilde{\ell}}^{[1]}  \Big(   \hat{S}^{+,\vec{\alpha}}_{\tilde{k},\tilde{ \ell}} \pm  \hat{S}^{-,\vec{\alpha}}_{\tilde{k},\tilde{ \ell}} \Big)    Y_{\tilde{k}, \tilde{\ell}}^{[1]}  \tilde{B}_{\tilde{k}, \tilde{\ell}} \Bigg( \rho_{\tilde{\ell}-1} \otimes \outerproduct{\mathbf{0
     }}{\mathbf{0}} \Bigg) \tilde{B}_{\tilde{k}, \tilde{\ell}}^{ \dagger}    \left( X_{\tilde{k}, \tilde{\ell}}^{[1]} Y_{\tilde{k}, \tilde{\ell}}^{[1]} \right)^{ \dagger} \tilde{A}_{\tilde{k}, \tilde{\ell}}^{ \dagger} \Bigg( \mathds{1} \otimes \sigma_{\tilde{\ell}} \Bigg) \Bigg) \\
     &-\Tr \Bigg( \Bigg( \mathds{1} \otimes \sigma_{\tilde{\ell}} \Bigg) \tilde{A}_{\tilde{k}, \tilde{\ell}}  X_{\tilde{k}, \tilde{\ell}}^{[1]}  Y_{\tilde{k}, \tilde{\ell}}^{[1]}  \tilde{B}_{\tilde{k}, \tilde{\ell}} \Bigg( \rho_{\tilde{\ell}-1} \otimes \outerproduct{\mathbf{0
     }}{\mathbf{0}} \Bigg) \tilde{B}_{\tilde{k}, \tilde{\ell}}^{ \dagger}  \left( Y_{\tilde{k}, \tilde{\ell}}^{[1]} \right)^{ \dagger}  \Big(   \hat{S}^{+,\vec{\alpha}}_{\tilde{k},\tilde{ \ell}} \pm  \hat{S}^{-,\vec{\alpha}}_{\tilde{k},\tilde{ \ell}} \Big)  \left( X_{\tilde{k}, \tilde{\ell}}^{[1]} \right)^{ \dagger} \tilde{A}_{\tilde{k}, \tilde{\ell}}^{ \dagger}  \Bigg) \\
     &+\Tr \Bigg( \tilde{A}_{\tilde{k}, \tilde{\ell}}  X_{\tilde{k}, \tilde{\ell}}^{[2]}  \Big(   \hat{S}^{+,\vec{\alpha}}_{\tilde{k},\tilde{ \ell}} \pm  \hat{S}^{-,\vec{\alpha}}_{\tilde{k},\tilde{ \ell}} \Big)    Y_{\tilde{k}, \tilde{\ell}}^{[2]}  \tilde{B}_{\tilde{k}, \tilde{\ell}} \Bigg( \rho_{\tilde{\ell}-1} \otimes \outerproduct{\mathbf{0
     }}{\mathbf{0}} \Bigg) \tilde{B}_{\tilde{k}, \tilde{\ell}}^{ \dagger}    \left( X_{\tilde{k}, \tilde{\ell}}^{[2]} Y_{\tilde{k}, \tilde{\ell}}^{[2]} \right)^{ \dagger} \tilde{A}_{\tilde{k}, \tilde{\ell}}^{ \dagger} \Bigg( \mathds{1} \otimes \sigma_{\tilde{\ell}} \Bigg) \Bigg) \\
     &-\Tr \Bigg( \Bigg( \mathds{1} \otimes \sigma_{\tilde{\ell}} \Bigg) \tilde{A}_{\tilde{k}, \tilde{\ell}}  X_{\tilde{k}, \tilde{\ell}}^{[2]}  Y_{\tilde{k}, \tilde{\ell}}^{[2]}  \tilde{B}_{\tilde{k}, \tilde{\ell}} \Bigg( \rho_{\tilde{\ell}-1} \otimes \outerproduct{\mathbf{0
     }}{\mathbf{0}} \Bigg) \tilde{B}_{\tilde{k}, \tilde{\ell}}^{ \dagger}  \left( Y_{\tilde{k}, \tilde{\ell}}^{[2]} \right)^{ \dagger}  \Big(   \hat{S}^{+,\vec{\alpha}}_{\tilde{k},\tilde{ \ell}} \pm  \hat{S}^{-,\vec{\alpha}}_{\tilde{k},\tilde{ \ell}} \Big)  \left( X_{\tilde{k}, \tilde{\ell}}^{[2]} \right)^{ \dagger} \tilde{A}_{\tilde{k}, \tilde{\ell}}^{ \dagger}  \Bigg) \\
     =& \Tr \Bigg( \tilde{A}_{\tilde{k}, \tilde{\ell}} \left( X_{\tilde{k}, \tilde{\ell}}^{[1]}  \Big(   \hat{S}^{+,\vec{\alpha}}_{\tilde{k},\tilde{ \ell}} \pm  \hat{S}^{-,\vec{\alpha}}_{\tilde{k},\tilde{ \ell}} \Big)    Y_{\tilde{k}, \tilde{\ell}}^{[1]} + X_{\tilde{k}, \tilde{\ell}}^{[2]}  \Big(   \hat{S}^{+,\vec{\alpha}}_{\tilde{k},\tilde{ \ell}} \pm  \hat{S}^{-,\vec{\alpha}}_{\tilde{k},\tilde{ \ell}} \Big)    Y_{\tilde{k}, \tilde{\ell}}^{[2]} \right) \tilde{B}_{\tilde{k}, \tilde{\ell}} \Bigg( \rho_{\tilde{\ell}-1} \otimes \outerproduct{\mathbf{0
     }}{\mathbf{0}} \Bigg) \mathcal{G}_{\tilde{\ell}}^\dagger \Bigg( \mathds{1} \otimes \sigma_{\tilde{\ell}} \Bigg) \Bigg) \\
     &- \Tr \Bigg( \Bigg( \mathds{1} \otimes \sigma_{\tilde{\ell}} \Bigg) \mathcal{G}_{\tilde{\ell}}   \Bigg( \rho_{\tilde{\ell}-1} \otimes \outerproduct{\mathbf{0
     }}{\mathbf{0}} \Bigg) \tilde{B}_{\tilde{k}, \tilde{\ell}}^{ \dagger} \left(  X_{\tilde{k}, \tilde{\ell}}^{[1]} \Big(   \hat{S}^{-,\vec{\alpha}}_{\tilde{k},\tilde{ \ell}} \pm  \hat{S}^{+,\vec{\alpha}}_{\tilde{k},\tilde{ \ell}} \Big)  Y_{\tilde{k}, \tilde{\ell}}^{[1]}  + X_{\tilde{k}, \tilde{\ell}}^{[2]}     \Big(   \hat{S}^{-,\vec{\alpha}}_{\tilde{k},\tilde{ \ell}} \pm  \hat{S}^{+,\vec{\alpha}}_{\tilde{k},\tilde{ \ell}} \Big)  Y_{\tilde{k}, \tilde{\ell}}^{[2]}  \right)^\dagger \tilde{A}_{\tilde{k}, \tilde{\ell}}^{ \dagger}  \Bigg) \\
     =& v_{\tilde{k}, \tilde{ \ell}}^{\vec{\alpha}} \mp \left( v_{\tilde{k}, \tilde{ \ell}}^{\vec{\alpha}} \right)^*  = \begin{cases} 2 i \Im \left(v_{\tilde{k}, \tilde{ \ell}}^{\vec{\alpha}} \right)  &, \quad \text{if} \quad + \\ 
     2 \Re \left(v_{\tilde{k}, \tilde{ \ell}}^{\vec{\alpha}} \right)  &, \quad \text{if} \quad -
     \end{cases}
\end{align*}
where 
\begin{align*}
    v_{\tilde{k}, \tilde{ \ell}}^{\vec{\alpha}}  =&  \Tr \Bigg( \tilde{A}_{\tilde{k}, \tilde{\ell}} \left( X_{\tilde{k}, \tilde{\ell}}^{[1]}  \Big(   \hat{S}^{+,\vec{\alpha}}_{\tilde{k},\tilde{ \ell}} \pm  \hat{S}^{-,\vec{\alpha}}_{\tilde{k},\tilde{ \ell}} \Big)    Y_{\tilde{k}, \tilde{\ell}}^{[1]} + X_{\tilde{k}, \tilde{\ell}}^{[2]}  \Big(   \hat{S}^{+,\vec{\alpha}}_{\tilde{k},\tilde{ \ell}} \pm  \hat{S}^{-,\vec{\alpha}}_{\tilde{k},\tilde{ \ell}} \Big)    Y_{\tilde{k}, \tilde{\ell}}^{[2]} \right) \tilde{B}_{\tilde{k}, \tilde{\ell}} \Bigg( \rho_{\tilde{\ell}-1} \otimes \outerproduct{\mathbf{0
     }}{\mathbf{0}} \Bigg) \mathcal{G}_{\tilde{\ell}}^\dagger \Bigg( \mathds{1} \otimes \sigma_{\tilde{\ell}} \Bigg) \Bigg).
\end{align*}
 \begin{figure*}[t]
\centering
    \includegraphics[width=0.8\textwidth]{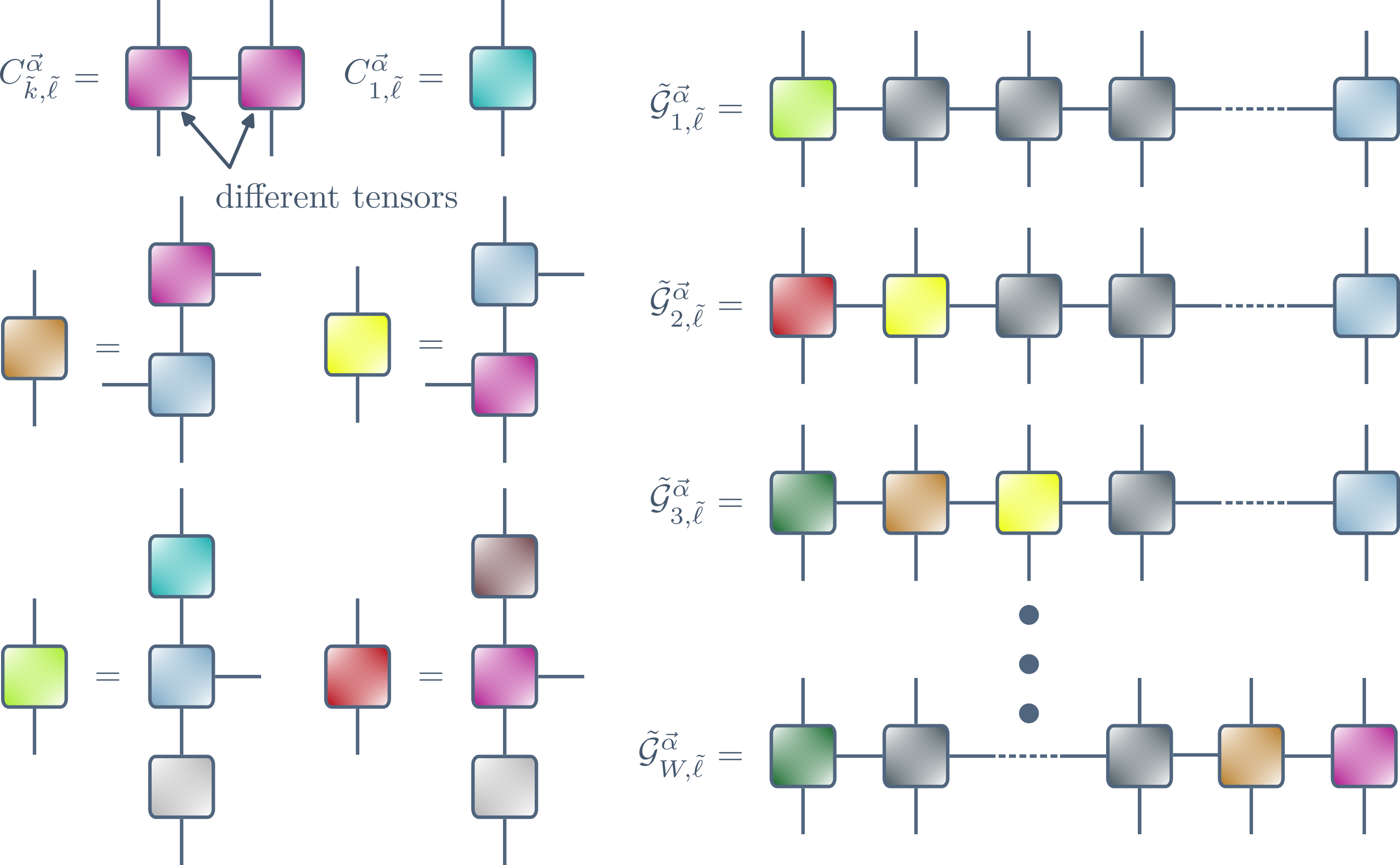}
    \caption{\textbf{MPOs for the update.} The gate $\tilde{\mathcal{G}}_{\tilde{k},\tilde{\ell}}^{\vec{\alpha}}$ differs from $\mathcal{G}_{\tilde{\ell}}$ by modification of $G_{\tilde{k},\tilde{\ell}}$ for $\tilde{k} \geq 2$ and $R_{1,\tilde{\ell}}$ for $\tilde{k}=1$. The modified gates are denoted $C_{\tilde{k},\tilde{\ell}}^{\vec{\alpha}}$. Writing also these operators as MPOs, we illustrate the possible forms for the MPO $\tilde{\mathcal{G}}_{\tilde{k},\tilde{\ell}}^{\vec{\alpha}}$ , based on Fig.~\ref{Fig1_sup}.
    }
    \label{Fig2_sup}
\end{figure*}
Now note that we can write both $u_{\tilde{k}, \tilde{ \ell}}^{\vec{\alpha}}$ and $v_{\tilde{k}, \tilde{ \ell}}^{\vec{\alpha}}$ as 
\begin{align*}
    \Tr_{ \tilde{ \ell}} \left( \Tr_{ \tilde{ \ell} - 1} \left( \tilde{A}_{\tilde{k}, \tilde{ \ell}} C_{\tilde{k}, \tilde{ \ell}}^{\vec{\alpha}} \tilde{B}_{\tilde{k}, \tilde{ \ell}}  \Bigg[ \rho_{\tilde{ \ell} - 1} \otimes  \ket{\mathbf{0}}\bra{\mathbf{0}} \Bigg] \mathcal{G}_{ \tilde{ \ell}}^{ \dagger} \right) \sigma_{\tilde{ \ell}} \right) .
\end{align*}
Here we defined 
\begin{align*}
C_{\tilde{k}, \tilde{ \ell}}^{\vec{\alpha}} &=  X_{\tilde{k}, \tilde{\ell}}^{[3]}  \hat{S}^{\vec{\alpha}}_{\tilde{k},\tilde{ \ell}}    Y_{\tilde{k}, \tilde{\ell}}^{[3]} \quad , \quad \text{for} \quad u_{\tilde{k}, \tilde{ \ell}}^{\vec{\alpha}}\\
    C_{\tilde{k}, \tilde{ \ell}}^{\vec{\alpha}} &=  \left( X_{\tilde{k}, \tilde{\ell}}^{[1]}  \Big(   \hat{S}^{+,\vec{\alpha}}_{\tilde{k},\tilde{ \ell}} \pm  \hat{S}^{-,\vec{\alpha}}_{\tilde{k},\tilde{ \ell}} \Big)    Y_{\tilde{k}, \tilde{\ell}}^{[1]} + X_{\tilde{k}, \tilde{\ell}}^{[2]}  \Big(   \hat{S}^{+,\vec{\alpha}}_{\tilde{k},\tilde{ \ell}} \pm  \hat{S}^{-,\vec{\alpha}}_{\tilde{k},\tilde{ \ell}} \Big)    Y_{\tilde{k}, \tilde{\ell}}^{[2]} \right) \quad , \quad \text{for} \quad v_{\tilde{k}, \tilde{ \ell}}^{\vec{\alpha}}
\end{align*}
or concretely
\begin{align*}
    C_{\tilde{k}, \tilde{ \ell}}^{\vec{\alpha}} &=  \left[ \chi_{\tilde{k}\geq 2} G_{\tilde{k},\tilde{\ell}} + \chi_{\tilde{k} = 1} R_{1,\tilde{\ell}}  \right] \hat{S}^{\vec{\alpha}}_{\tilde{k},\tilde{ \ell}} \quad , \quad \text{for} \quad u_{\tilde{k}, \tilde{ \ell}}^{\vec{\alpha}} \\
    C_{\tilde{k}, \tilde{ \ell}}^{\vec{\alpha}}&= T_{\tilde{k}, \tilde{ \ell}} \bigg(  \Big(   \hat{S}^{+,\vec{\alpha}}_{\tilde{k},\tilde{ \ell}} \pm  \hat{S}^{-,\vec{\alpha}}_{\tilde{k},\tilde{ \ell}} \Big)   e^{-i \sqrt{\delta t} V_{\tilde{k}, \tilde{ \ell}}} + e^{-i \sqrt{\delta t} V_{\tilde{k}, \tilde{ \ell}}} \Big(   \hat{S}^{+,\vec{\alpha}}_{\tilde{k},\tilde{ \ell}} \pm  \hat{S}^{-,\vec{\alpha}}_{\tilde{k},\tilde{ \ell}} \Big)     \bigg) e^{-i \delta t H_{\tilde{k}, \tilde{ \ell}}} \quad , \quad \text{for} \quad v_{\tilde{k}, \tilde{ \ell}}^{\vec{\alpha}}.
\end{align*}
The inner trace can be straightforwardly translated into doubled space via (1)' by defining $ \tilde{\mathcal{G}}_{\tilde{k},\tilde{\ell}}^{\vec{\alpha}} = \tilde{A}_{\tilde{k}, \tilde{\ell}} C_{\tilde{k}, \tilde{\ell}}^{\vec{\alpha}} \tilde{B}_{\tilde{k}, \tilde{\ell}}$
\begin{equation*}
    \ket{\tilde{\rho}_{\tilde{\ell}}} = \bra{\mathds{1}} \left( \tilde{\mathcal{G}}_{\tilde{k},\tilde{\ell}}^{\vec{\alpha}} \otimes \mathcal{G}_{\tilde{\ell}}^* \right) \left( \mathds{1} \otimes \ket{\mathbf{0}} \otimes \mathds{1} \otimes \ket{\mathbf{0}} \right) \ket{\rho_{\tilde{\ell}-1}} .
\end{equation*}
In Fig.~\ref{Fig2_sup} it is shown how the MPOs for $\tilde{\mathcal{G}}_{\tilde{k},\tilde{\ell}}^{\vec{\alpha}}$ can be obtained. As a last step, we need to express $\Tr(\tilde{\rho}_{\tilde{\ell}} \sigma_{\tilde{\ell}})$ in doubled-space. We have 
\begin{align*}
    \Tr(\tilde{\rho}_{\tilde{\ell}} \sigma_{\tilde{\ell}}) &= \bra{\tilde{\rho}_{\tilde{\ell}}}  \ket{\sigma_{\tilde{\ell}}} ,
\end{align*}
which is just an overlap between the two (matrix-product) states.  
\subsection{Tensor network parameters}
 \begin{figure*}[t]
\centering
    \includegraphics[width=\textwidth]{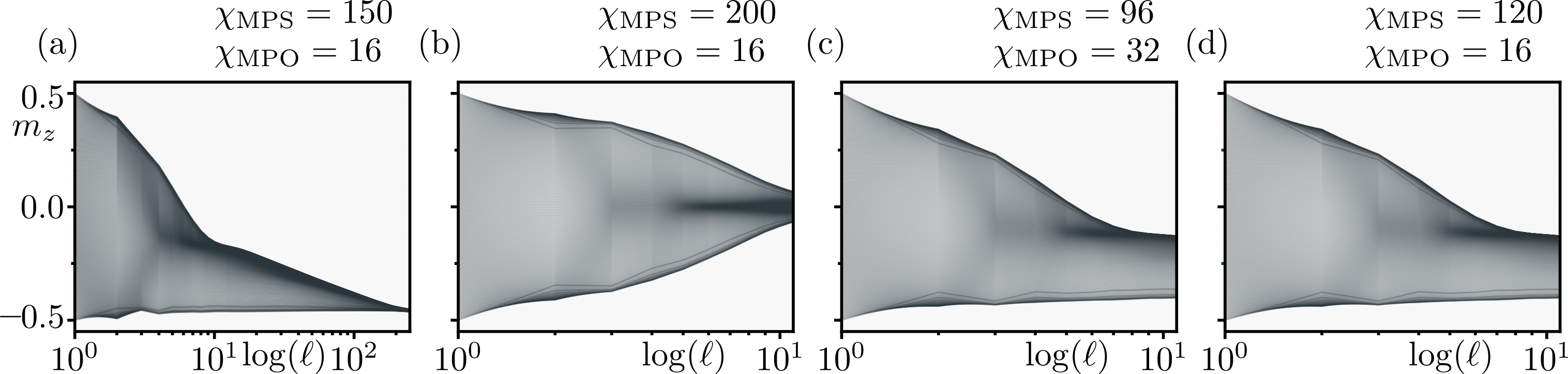}
    \caption{\textbf{Checks on numerical data.} (a) Evolution of the magnetization for a QNN with $W=40$. The local gates are chosen as in Fig.~\ref{Fig3}(b) but the MPS bond dimension for simulation is higher ($\chi_{\mathrm{MPS}}=150$). (b) For the initial QNN with $W=20$ [cf. Fig.~\ref{Fig3}(a)] we have chosen a substantially higher MPS bond dimension $\chi_{\mathrm{MPS}}=200$ and observed  that the dynamical behavior of the magnetization largely agrees with the one in Fig.~\ref{Fig3}(a). (c) Evolution of the desired QNN with MPS bond dimension $\chi_{\mathrm{MPS}}=96$ and MPO bond dimension $\chi_{\mathrm{MPO}}=32$. (d) Evolution of the desired QNN with MPS bond dimension $\chi_{\mathrm{MPS}}=120$ and MPO bond dimension $\chi_{\mathrm{MPO}}=16$. Comparing (c) and (d), one finds that neither increasing $\chi_{\mathrm{MPS}}$ nor increasing $\chi_{\mathrm{MPO}}$ seem to have an effect on the evolution of the magnetization. Further notice the similarity between the evolution of the desired QNN in panels (c) and (d) and the one of the trained QNN in Fig.~\ref{Fig3}(d).}
    \label{Fig3_sup}
\end{figure*}
Here we provide details on parameters used for data generation. Note that the time evolutions for the magnetization in Fig.~\ref{Fig2}(a) can be obtained without resorting to MPOs/MPSs since for the small width $W=4$ the input-output relation (forward channel) of the QNN can be simulated exactly. For $W=40$ [see Fig.~\ref{Fig2}(b)], on the other hand, we have to use tensor network methods. We used MPSs with bond dimension $\chi_{\mathrm{MPS}}=120$. We checked that this bond dimension is appropriate by validating data through test simulations for higher bond dimensions, for which the dynamical behavior of the magnetization shows no qualitative changes and only slight quantitative deviations [see Fig.~\ref{Fig3_sup}(a)]. Similarly, we compressed the MPO to a bond dimension $\chi_{\mathrm{MPO}}=16$, for which the square root of the sum of the squared discarded singular values (truncation error) is of the order of machine precision.\\
Due to the enormous resource requirements, for the training process we choose the smallest possible bond dimension for which the algorithm still yields faithful results. Thus, for training as analyzed in Fig.~\ref{Fig3} of the main text, we have chosen an MPS bond dimension $\chi_{\mathrm{MPS}}=96$ which is sufficient for the depths we are considering, where accumulation of truncation errors can be neglected. The MPO bond dimension is $\chi_{\mathrm{MPO}}=16$ and also here the truncation error is of the order of machine precision. Indeed, we verified in Fig.~\ref{Fig3_sup}(b-d) that for higher bond dimensions no qualitative and only slight quantitative changes can be observed in the evolutions of desired and initial QNNs. During the training, we monitored the MPS-truncation error, which was of the order of the one for the initial and desired QNNs.

\section{Metastability in open quantum systems}
Here we want to explain how metastability can occur in principle in Markovian open quantum dynamics \cite{macieszczak2016, rose2016, macieszczak2021}. For this, let the dynamics of an open $W$-spin system coupled to an environment be governed by a Markovian quantum dynamical semigroup, generated by the Lindbladian (see main text)
\begin{align*}
    \mathcal{L}[O] &= -i[H, O] + \sum_{k=1}^W \big( J_k O J_k^\dagger - \frac{1}{2} \{ J_k^\dagger J_k, O \} \big) ,
\end{align*}
such that states, represented by a density matrix $\rho_0$, evolve as $\rho_t = e^{t\mathcal{L}}[ \rho_0]$. The operator $H$ occurring in the coherent part is the Hamiltonian and the jump operators $J_k$ form the dissipative part. Since $\mathcal{L}$ is a linear operator, we can find its eigenvalues $\{ \lambda_k , k=1,\ldots, d^2 \}$, and sort them in decreasing order according to their real part $\Re(\lambda_k) \geq \Re(\lambda_{k+1})$. Since dynamics is trace-preserving and completely positive we have that $\lambda_1 = 0$ with left eigenmatrix $L_1=\mathds{1}$ and right eigenmatrix $\rho_{ss}$ (stationary state). If left and right eigenmatrices $\{ L_k \}, \{ R_k \}$ are complete biorthonormal bases we may expand
\begin{equation} \label{expansion}
    \rho_t = \rho_{ss} + \sum_{k=2}^{d^2} e^{t \lambda_k} c_k R_k ,
\end{equation}
with $c_k = \Tr(L_k \rho_0)$. Now assume that we have a separation in the spectrum $|\Re(\lambda_m)| \ll |\Re(\lambda_{m+1})|$. This defines the two time-scales $\tau = -1/\Re(\lambda_m)$ and $\tau' = - 1 / \Re(\lambda_{m+1})$ with $\tau' \ll \tau$. In the intermediate time-regime $\tau' \ll t \ll \tau$ we have that $|e^{t \Re(\lambda_k)}| \approx 1$ for $2 \leq k \leq m$ and $e^{t \Re\lambda_k)} \approx 0$ for $m+1 \leq k \leq d^2$ such that Eq.~\eqref{expansion} becomes 
\begin{equation} 
    \rho_t \approx \rho_{ss} + \sum_{k=2}^{m}  e^{it\Im(\lambda_k)}c_k R_k .
\end{equation}
In cases in which all the eigenvalues $\lambda_k$, with $k\le m$, are real (which is for instance what happens for the considered dissipative quantum Ising model), the state $\rho_t$ becomes approximately stationary  and depends on $\rho_0$, i.e. ergodicity is partially broken. However, on even longer timescales  the state is eventually driven towards $\rho_{ss}$.

\end{document}